\documentclass[12pt]{article}
\usepackage{graphicx} 
\usepackage{bm}
\usepackage{amsmath}

\usepackage[authoryear]{natbib} 

\usepackage{float} 
\usepackage{booktabs} 
\usepackage{multirow} 
\usepackage{subfig} 
\usepackage[left=1in, right=1in, top=1in, bottom=1in]{geometry}
\usepackage{appendix} 
\usepackage{amssymb} 
\usepackage[affil-it]{authblk} 

\PassOptionsToPackage{hyphens}{url} 
\usepackage{hyperref} 
\hypersetup{
    colorlinks=true,
    citecolor=blue,
    linkcolor=blue,
    filecolor=magenta,      
    urlcolor=cyan,
}

\usepackage[dvipsnames]{xcolor}
\definecolor{gray}{rgb}{0.5,0.5,0.5}
\usepackage{listings}
\lstdefinelanguage{rstan}{
morekeywords={data,int,matrix,vector,real,lower,transformed,parameters,for,in,target,generated,quantities,model,rock,around,the,tonight},
morecomment=[l]{//}
}
\title{A spatio-temporal model to detect potential outliers in disease mapping}
\author[1]{Victoire Michal\thanks{Corresponding author: victoire.michal@mail.mcgill.ca}}
\author[1]{Alexandra M. Schmidt}
\affil[1]{\small Department of Epidemiology, Biostatistics and Occupational Health, McGill University, Montreal, Canada}
\date{}

\begin{document}
\fontsize{10pt}{20pt}\selectfont

\maketitle

\begin{abstract}
Spatio-temporal disease mapping models are commonly used to estimate the relative risk of a disease over time and across areas. For each area and time point, the disease count is modelled with a Poisson distribution whose mean is the product of an offset and the disease relative risk. This relative risk is commonly decomposed in the log scale as the sum of fixed and latent effects. The Rushworth model allows for spatio-temporal autocorrelation of the random effects. We build on the Rushworth model to accommodate and identify potentially outlying areas with respect to their disease relative risk evolution, after taking into account the fixed effects. An area may display outlying behaviour at some points in time but not all. At each time point, we assume the latent effects to be spatially structured and include scaling parameters in the precision matrix, to allow for heavy-tails. Two prior specifications are considered for the scaling parameters: one where they are independent across space and one with spatial autocorrelation. We investigate the performance of the different prior specifications of the proposed model through simulation studies and analyse the weekly evolution of the number of COVID-19 cases across the 33 boroughs of Montreal and the 96 French departments during the second wave. In Montreal, 6 boroughs are found to be potentially outlying. In France, the model with spatially structured scaling parameters identified 21 departments as potential outliers. We find that these departments tend to be close to each other and within common French regions.
\end{abstract}



\section{Introduction}

Since 1995, disease mapping models have been proposed to estimate the spatio-temporal trend of relative risks \citep{Bernardinelli, Lawson_Book}. In the literature on spatio-temporal disease mapping, models commonly assume that the disease cases follow a Poisson distribution whose mean is the product between an offset and the relative risk, which varies through time and across space. The relative risk is usually written as the sum of fixed and latent effects. Commonly, the random effects are spatially structured and evolve through time (see, e.g., \cite{CARBayesST} for an overview). Further, in their discussion, \cite{Rushworth} note that two neighbouring areas may behave differently over time, which is usually not accounted for in spatio-temporal disease mapping models. We propose a spatio-temporal model that identifies potential outliers with respect to the disease risk. In particular, the proposed model aims to identify any area whose behaviour over time differs from their neighbours or the rest of the region of interest. 
This provides decision makers with tools to help prioritise interventions and implement localised policies.

\cite{Knorr-held} proposed a spatio-temporal disease mapping model wherein the log relative risks are decomposed as the sum of fixed, temporal, spatial and space-time interaction effects. Four different interaction patterns are considered, the more complex case assuming interaction terms that are both spatially and temporally structured. The covariance matrix for this interaction term is constructed following \cite{clayton1996}. It is given by the Kronecker product between a temporal random walk structure and a spatial intrinsic conditional autoregressive (ICAR) structure \citep{CAR}. \cite{ugarte2012} built on \cite{Knorr-held} to model the spatial effects following a Leroux prior \citep{Leroux}, while keeping the covariance matrix for the space-time interaction term as the combination of a random walk and ICAR interaction structure. \cite{Rushworth} proposed to reduce the parameter space by including only the space-time interaction effects and further introduced the Leroux structure in the spatio-temporal structure. More specifically, let $\bm{b}_{\cdot t} = [b_{1t}, \dots, b_{nt}]^\top$ be the vector of areal latent effects at time $t$, they assume 
\begin{equation}
    \label{eq:Rush}
    \bm{b}_{\cdot 1} \sim \mathcal{N}\left(\bm{0}, \sigma^2\bm{Q}_L^{-1}\right), \quad \mbox{and} \quad \bm{b}_{\cdot t} \mid \bm{b}_{\cdot t-1} \sim \mathcal{N}\left(\alpha\bm{b}_{\cdot t-1},\sigma^2\bm{Q}_L^{-1} \right), \ t=2, \dots, T,
\end{equation}
with conditional variance $\sigma^2$, temporal smoothing parameter $\alpha \in [0,1]$ and spatial smoothing parameter $\lambda \in [0,1)$. The precision matrix $\bm{Q}_L = (1-\lambda)\bm{I} + \lambda(\bm{D}-\bm{W})$ is the one proposed by \cite{Leroux}, with $\bm{W}=[w_{ij}]$ a $n \times n$ matrix of spatial weights and $\bm{D}=diag(d_i)$, for $d_i = \sum_{j \neq i} w_{ij}$. The spatial weights are commonly defined as $w_{ij}=1$ if areas $i$ and $j$ share a border, and $w_{ij}=0$ otherwise. Hence, \cite{Rushworth} assume a non-separable spatio-temporal structure, where the temporal trend appears in the conditional means and the spatial structure, in the precision. 

On the other hand, the models mentioned above assume spatial homogeneity through time. 
In the purely spatial setting, this issue of spatial heterogeneity may be addressed by allowing the spatial structure $\bm{W}$ to be estimated from the data (see, e.g., \cite{lee2013_estW, dean2019frontiers, corpasburgos}). Another approach is to allow the spatial dependence parameter $\lambda$ to vary across space, which accommodates local differences in the spatial structure \citep{macnab2023}. Other authors proposed two-step procedures to elicit clusters of areas that behave similarly and include that information in the spatial model \citep{clust_anderson2014, clust_santafe2021}. \cite{Congdon} allowed for disparities by modifying the Leroux prior to include scaling mixture components $\kappa_i>0, i=1, \dots, n,$ such that $\bm{b}=[b_1, \dots, b_n]^\top \sim \mathcal{N}(\bm{0}, \sigma^2\bm{Q}_C^-),$ where the precision matrix has diagonal elements $\bm{Q}_{C, ii} = \kappa_i(1-\lambda + \lambda d_i)$ and off-diagonal elements $\bm{Q}_{C, ij} = -\lambda w_{ij}\kappa_i\kappa_j$. This joint distribution proposed by \cite{Congdon} corresponds to the following conditional distributions:
\begin{equation}
    \label{eq:Congdon}
    b_i \mid \bm{b}_{(-i)} \sim \mathcal{N}\left(\frac{\lambda}{1-\lambda + \lambda d_i}\sum_{j=1}^n w_{ij}\kappa_jb_j, \frac{\sigma^2}{\kappa_i\left(1-\lambda +\lambda d_i\right)}\right), \ i=1, \dots, n.
\end{equation}
In this proposal, the scaling mixture parameters help identify potential outliers, wherein $\kappa_i<1$ implies that the $i$th area is a potential outlier. Let area $i$ be an outlier and a neighbour of area $j$. Then $\kappa_i<1$ inflates the conditional variance for $b_i$ while allowing $b_j$ to allocate less weight to the outlying $b_i$ in its conditional mean, and borrow more strength from its non-outlying neighbours.

In the spatio-temporal setting, proposals have been made to extend some methods discussed in the previous paragraph. For instance, \cite{spatiotemp_clust_lee2016} proposed to include a piecewise constant intercept term to identify clusters of areas that behave similarly over space and time. \cite{rushworth2017} proposed a spatio-temporal model where the spatial structure is estimated based on the data. Different from these methods, the main aim of this paper is to propose a spatio-temporal model that accommodates and specifically identifies potential outlying areas, after accounting for fixed effects. Specifically, we propose to extend Congdon's prior in equation  (\ref{eq:Congdon}) to the spatio-temporal setting, similarly to how \cite{Rushworth} (\ref{eq:Rush}) extended the Leroux prior \citep{Leroux}. Throughout this paper, the term outlier designates both areas that may behave differently from their neighbours (spatial outliers), and areas that present extreme risks.

\subsection{Illustration}
\label{sec:Motivation}

\begin{figure}[H]
    \centering
    \includegraphics[width=\textwidth]{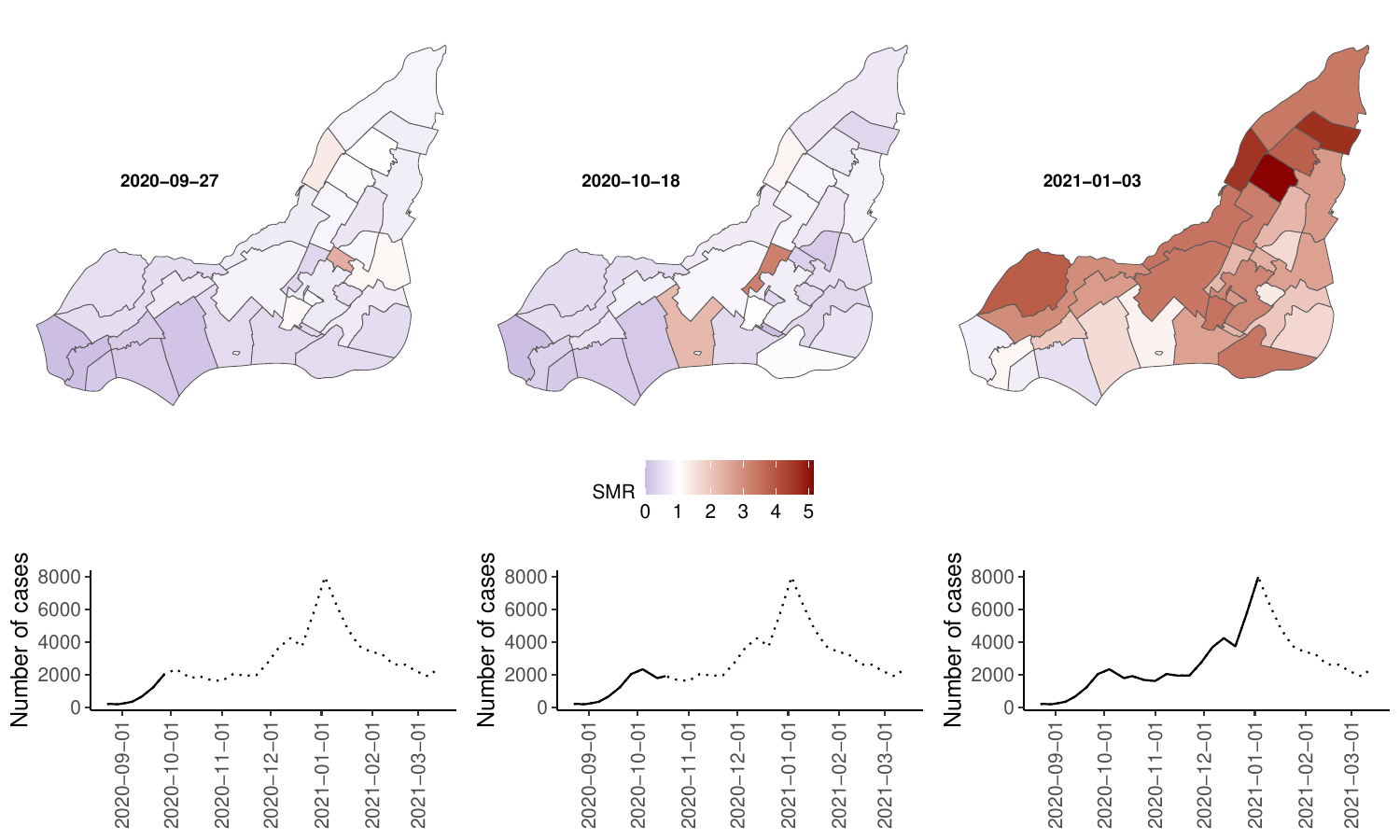}
    \caption{\footnotesize Maps of the SMR distribution across the boroughs of Montreal at three different time points (top) and distribution of the total number of COVID-19 cases over time (bottom).}
    \label{fig:Covid_MapsSMR_NumCases}
\end{figure}

\begin{figure}[H]
    \centering
    \includegraphics[width=\textwidth]{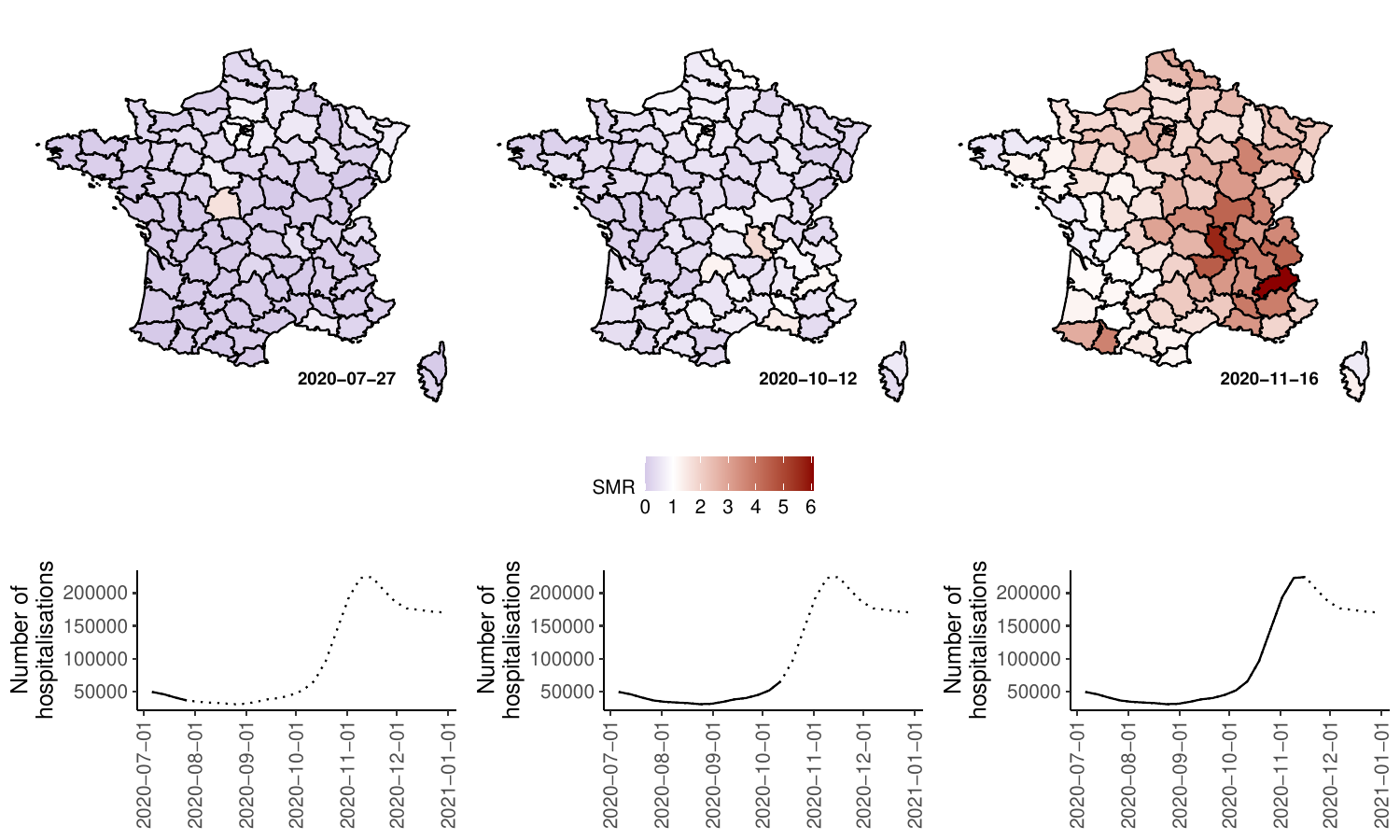}
    \caption{\footnotesize Maps of the SMR distribution across the French departments at three different time points (top) and distribution of the total number of COVID-19 hospitalisations over time (bottom).}
    \label{fig:Covid_MapsSMR_NumCases_Fr}
\end{figure}

To investigate the benefits of the proposed model, we consider two examples related to the coronavirus disease 2019 (COVID-19) pandemic. 
Due to the spatial dimension of the disease, disease mapping and spatio-temporal methods have been widely used to analyse COVID-19 counts (see, e.g., \cite{Covid_spatial} for a review) in order to help decision makers understand the disease and implement policies. Further, COVID-19 counts tend to show different behaviours over time and across areas (see, e.g., Figures \ref{fig:Covid_MapsSMR_NumCases} and \ref{fig:Covid_MapsSMR_NumCases_Fr} for the behaviour of COVID-19 standardised morbidity ratios (SMRs) during the second wave in Montreal and in France).

First, we have data available across the 33 boroughs of Montreal, where the number of COVID-19 cases have been recorded weekly during the second wave, between August 23rd 2020 and March 20th 2021 (see, e.g., \cite{INSPQ_CovidTimeline} for a COVID-19 timeline in the province of Quebec). The data come from the \textit{Institut national de la santé publique du Québec} (INSPQ). 
Figure \ref{fig:Covid_MapsSMR_NumCases} showcases the SMR distribution for the COVID-19 cases across space at three different time points. Even with limiting policies in place \citep{INSPQ_CovidTimeline}, some boroughs appear to have elevated SMRs at some time points but not all, while other boroughs seem to never show extreme values. Similar to \cite{michal2022joint}, three auxiliary variables are considered to analyse these weekly COVID-19 cases, namely the number of beds in long-term care centres (\textit{Centres d’hébergement et de soins de longue durée}, CHSLDs), the median age by borough and the population aged 25-64 with a university degree \citep{Mtl_age_diplome}.

As another example of the need to investigate outlying observations in spatio-temporal disease counts, we study the COVID-19 second wave in France. 
We have available the weekly counts of hospitalisation due to COVID-19 during the second wave, across the 96 French departments. In France, the second wave lasted 26 weeks between early July 2020, and the end of the year 2020 \citep{INSEE_CovidTimeline}. The data are publicly available from the French national health agency \citep{Covid_Fr_Data}. Figure \ref{fig:Covid_MapsSMR_NumCases_Fr} shows the evolution of the COVID-19 SMRs across the French departments. It appears that some departments might behave differently than the others, in particular at the beginning of the second wave. 

This paper is organised as follows. Section \ref{sec:Methods} proposes the spatio-temporal model that accounts for and identifies outlying areas over time. Therein, the inference procedure, which is performed following the Bayesian paradigm, is also discussed. Section \ref{sec:Analyses} shows the performance of the proposed model under different simulation scenarios (Section \ref{sec:Simulation}). Then, the proposed model is fitted to the COVID-19 data in Montreal (Section \ref{sec:Covid_Mtl}) and in France (Section \ref{sec:Covid_Fr}). Section \ref{sec:Discussion} provides concluding remarks and points to potential future avenues of research.

\section{Proposed model}
\label{sec:Methods}

Consider a region divided into $n$ non-overlapping areas studied over $T$ time points. Let $Y_{it}$ be the number of cases of a disease recorded in area $i$ at time $t$. Let $E_i$ be the expected number at risk in area $i$, which we assume to be constant over time. The number of cases is modelled as follows:
$$Y_{it} \mid E_i, \mu_{it} \sim \mathrm{Pois}\left(E_i \mu_{it}\right),$$
where $\mu_{it}$ is the relative disease risk for the $i$th area at time $t$, which is decomposed as
$$\log\left(\mu_{it}\right) = \beta_0 + \bm{x}_i^\top\bm{\beta} + b_{it},$$
where the intercept $\beta_0$ corresponds to the overall log risk across time and space, the $p$ regression coefficients $\bm{\beta}$ multiply the vector of areal-level covariates $\bm{x}_i$, and $b_{it}$ is a latent effect for the $i$th area at time $t$, which captures whatever is left after accounting for the covariates. A square $n \times n$ matrix of weights $\bm{W}=[w_{ij}]$ is defined to account for a spatial structure in the latent effects. Two areas $i$ and $j$ are said to be neighbours with a weight $w_{ij}=1$ if they share a border and $w_{ij}=0$, otherwise. From this 0-1 neighbourhood structure, the diagonal matrix $\bm{D}=diag(d_i)$, where $d_i = \sum_j w_{ij}$, corresponds to the matrix whose diagonal elements are the areal numbers of neighbours.
We propose a modification of the Rushworth prior (\ref{eq:Rush}) to model the vector of latent effects $\bm{b} = \left[b_{11}, \dots, b_{n1}, \dots, b_{1T}, \dots, b_{nT}\right]^\top$. Let $\bm{b}_{\cdot t} = [b_{1t}, \dots, b_{nt}]^\top$, we assume
\begin{align}
\label{eq:HeavyRush}
    \bm{b}_{\cdot 1} \sim \mathcal{N}\left(\bm{0}, \sigma^2\bm{Q}_C^{-}\right), \quad \mbox{and} \quad \bm{b}_{\cdot t} \mid \bm{b}_{\cdot t-1} \sim \mathcal{N}\left(\alpha\bm{b}_{\cdot t-1},\sigma^2\bm{Q}_C^{-} \right), \ t=2, \dots, T,
\end{align}
with temporal dependence parameter $|\alpha| < 1$, variance parameter $\sigma>0$, and $\bm{Q}_C$ as proposed by \cite{Congdon}. The matrix $\bm{Q}_C$ has diagonal elements $\bm{Q}_{C_{ii}} = \kappa_i(1-\lambda + \lambda d_i)$ and off-diagonal elements $\bm{Q}_{C_{ij}} = -w_{ij}\lambda \kappa_i \kappa_j,$ where $\lambda \in [0,1]$ and $\kappa_i > 0, \ i=1,\ \dots n$. The matrix $\bm{Q}_C$ may be written as
\begin{equation}
\label{eq:PrecMatCongdon}
    \bm{Q}_C = diag_1(\bm{\kappa}) \odot \left[(1-\lambda)\bm{I} + \lambda \left(\bm{D}-\bm{W}\odot \bm{\kappa}\bm{\kappa}^\top\right)\right],
\end{equation}
where $diag_1(\bm{\kappa})$ denotes the square matrix with diagonal elements $\bm{\kappa}=[\kappa_1, \dots, \kappa_n]^\top$ and off-diagonal elements equal to 1. 
From expression (\ref{eq:PrecMatCongdon}), it is clear that the mixing parameter $\lambda$ is a spatial dependence parameter and $\lambda=0$ yields temporal latent effects that are independent across space, while $\lambda=1$ implies fully structured spatio-temporal effects. Similarly, $\alpha$ appears as a temporal dependence parameter in the prior distribution (\ref{eq:HeavyRush}), where $\alpha=0$ leads to vectors of latent effects $\bm{b}_{\cdot t}$ that are spatially structured and independent over time, and $\alpha=1$ implies fully structured spatio-temporal effects. 

Our main contribution lies in the inclusion of the scaling parameters $\bm{\kappa}$. 
In expression (\ref{eq:PrecMatCongdon}), they appear in the diagonal elements of the precision matrix, and $\kappa_i<1$ inflates the conditional variance of the latent effects for the $i$th area. Additionally, the $\kappa$'s impact the spatial weights as follows: $\bm{W}\odot \bm{\kappa}\bm{\kappa}^\top = [w_{ij}\kappa_i\kappa_j]$. Hence, at any time $t$, $\kappa_i<1$ implies an inflated conditional variance for the $i$th latent effect, and a decreased correlation between areas $j$ and $i$ when they are neighbours. Following \cite{Congdon}, these parameters act as outlier indicators and an area $i$ is defined as a potential outlier if $\kappa_i<1$. An area may be outlying at all time points, or at some points in time.

Further, the role of the scaling parameters can be studied from the conditional distributions of the latent effects that result from (\ref{eq:HeavyRush}). 
For $t \geq 2$, the joint distribution of the latent effects, (\ref{eq:HeavyRush}), corresponds to the set of $n$ Gaussian conditional distributions with expectation $\mathrm{E}\left(b_{it} \mid \bm{b}_{\cdot t-1}, \bm{b}_{(-i) t}\right) = \alpha b_{it-1} + \lambda/(1-\lambda + \lambda d_i)\sum_{j \sim i}\kappa_j(b_{jt}-\alpha b_{jt-1})$ and variance $\mathrm{V}\left(b_{it} \mid \bm{b}_{\cdot t-1}, \bm{b}_{(-i) t}\right) = \sigma^2/\left(\kappa_i(1-\lambda + \lambda d_i)\right),$ where $\bm{b}_{(-i) t} = [b_{1t}, \dots, b_{i-1 t}, b_{i+1 t}, \dots, b_{nt}]^\top$ and  $j \sim i$ means that areas $i$ and $j$ are neighbours.
Hence, for outlying area $j$ a neighbour of area $i$, $\kappa_j$ is smaller than 1 and the difference $b_{jt}-\alpha b_{j t-1}$ contributes less to the conditional mean of $b_{it}$ than another neighbour $\ell$ whose $\kappa_\ell$ is greater or equal to 1.

Two priors are proposed  for the scaling mixture components. 
First, following \cite{Congdon}, independent gamma priors are assigned to the scaling mixture components, $\kappa_i \overset{i.i.d.}{\sim} \mathrm{Gamma}(\nu/2, \nu/2)$. This implies that $\mathbb{E}(\kappa_i \mid \nu)=1$ and $\mathbb{V}(\kappa_i \mid \nu) = 2/\nu$, \textit{a priori}; that is, the prior assumption is that area $i$ is not an outlier, with small variance for large hyperparameter $\nu$. Following \cite{Gelman_t4, Michal_zika}, we assume $\nu \sim \mathrm{Exp}(1/4)$. Second, we investigate a discretisation of the continuous spatially structured scaling process proposed by \cite{PalaciosSteel}. For this discrete parametrisation of \cite{PalaciosSteel}, we follow \cite{Michal_zika}, who assign a proper conditional autoregressive (PCAR) prior to the scaling components. They assume $\ln(\kappa_i) \equiv - \nu/2 + z_i,$ with scaled spatially structured $\bm{z} \equiv [z_1, \dots, z_n]^\top \sim \mathcal{N}\left(\bm{0}, \nu \bm{Q}_{\rho, \star}^{-1}\right)$ and $\nu \sim \mathrm{Exp}(1/0.3).$ The precision matrix $\bm{Q}_{\rho} = \bm{D}-\rho\bm{W}$, which is positive definite for $\rho \in [0,1)$ \citep{Banerjee}, is scaled by $h_{\rho} = \exp\left[(1/n)\sum_{i=1}^n \ln\left(\bm{Q}_{\rho, ii}^{-1}\right)\right]$ such that $\bm{Q}_{\rho, \star} = h_{\rho}\bm{Q}_{\rho}$. This scaling of the spatially structured precision matrix yields approximate marginal variances $\mathbb{V}(\ln(\kappa_i) \mid \nu) \simeq \nu$, for any spatial structure under study \citep{BYM2}. Similar to the case of independent gamma priors on the scaling components, this spatially structured prior implies that $\mathbb{E}(\kappa_i\mid \nu) =1,$  which means that the $i$th area is not an outlier \textit{a priori}. For further discussion regarding these two priors for the $\kappa$ parameters, see \cite{Michal_zika}.

\subsection{Inference procedure}
\label{sec:inf_proc}

The proposal (\ref{eq:HeavyRush}) and the Rushworth model (\ref{eq:Rush}) do not yield posterior distributions with a closed form. Therefore, to approximate the posterior distribution of the resultant parameter vector we resort to Markov Chain Monte Carlo (MCMC) methods. Specifically, we use the \texttt{R} package \texttt{rstan} \citep{Stan}, which efficiently estimates posterior distributions from complex hierarchical models where spatial and temporal structures are studied, using a sampler based on a Hamiltonian Monte Carlo algorithm \citep{Morris}. 

To avoid a potential identifiability issue between the intercept and the latent effects in the proposed model \citep{RueHeld}, we impose a sum-to-zero constraint on $\bm{b}_{\cdot 1}$, the latent effects for the first time point. In the MCMC procedure, a soft sum-to-zero constraint corresponds to assuming $\sum_{i=1}^nb_{i1} \sim \mathcal{N}(0, 0.001n)$ \citep{Morris}. The \texttt{rstan} code implemented to fit the proposed model in the simulation studies and data applications summarised in Section \ref{sec:Analyses} is displayed in Appendix \ref{App:StanCode}. For more details on the data and code, see \url{https://github.com/vicmic13/SpatioTemporal_DiseaseMapping_OutlyingAreas}. 

\section{Data analyses}
\label{sec:Analyses}

In Section \ref{sec:Simulation}, we present the results from a simulation study where the goal is to assess the performance of the proposed model (\ref{eq:HeavyRush}) compared to the Rushworth model (\ref{eq:Rush}). Data are first generated from the Rushworth model and some areas are contaminated into outliers that we aim to identify. Note that in Appendix \ref{App:SimFromModel}, we present results from a simulation study where data are generated from the proposed model to check whether we can recover the parameters used to generate the data. The results suggest that we can estimate the parameters of the model and there does not seem to be any identifiability issue in the model. Finally, Sections \ref{sec:Covid_Mtl} and \ref{sec:Covid_Fr} provide the analyses of the COVID-19 data across the 33 boroughs of Montreal and the 96 French departments. 

\subsection{Simulation study}
\label{sec:Simulation}

The simulation study presented in this section is inspired by the analyses shown in Appendix C of \cite{fonseca2023} and by the simulation studies summarised in \cite{Michal_zika}. We aim to investigate the performance of the proposed model when compared to the Rushworth model, and assess its ability to identify outliers when the truth is known. The region of interest is Montreal, which is divided into $n=33$ boroughs, and the time period is arbitrarily set to $T=52$ time points, in order to mimic weekly data recorded over a year. A known set of boroughs is contaminated at some of the time points, but not all, into outlying areas. The goal is to identify these areas that sometimes present outlying risks. 

\begin{figure}[H]
    \centering
    \includegraphics[width=\textwidth]{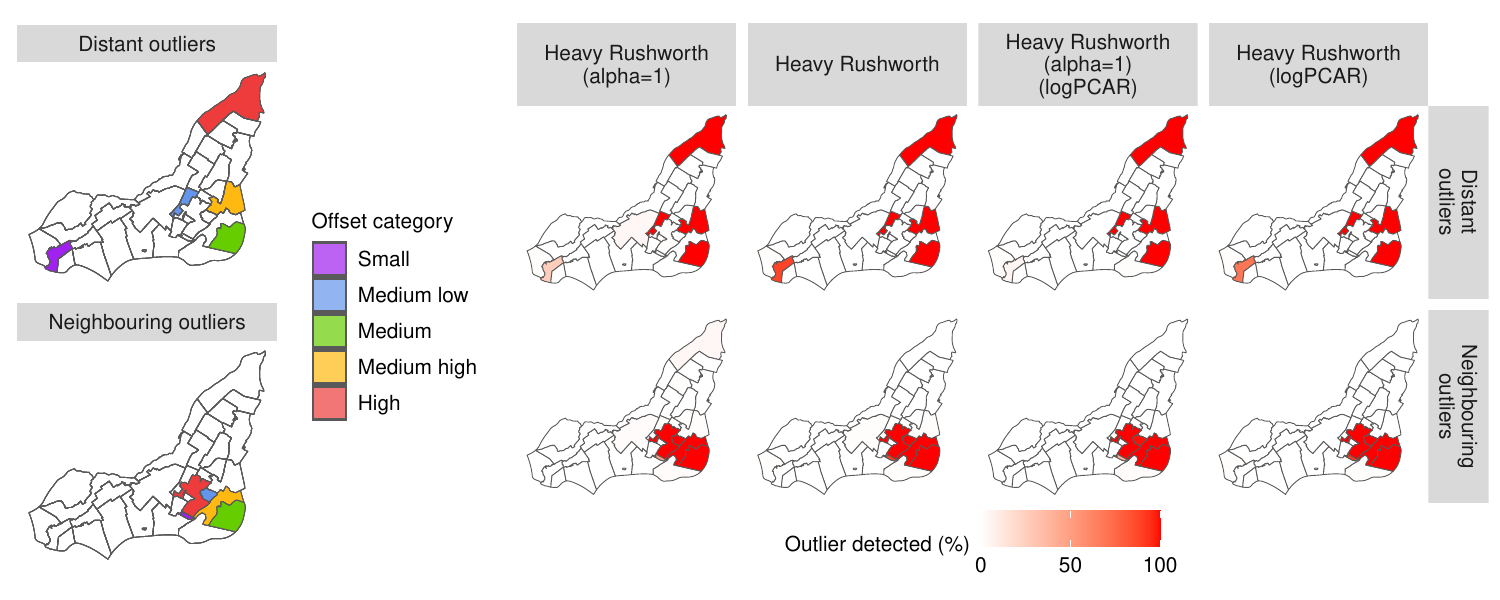}
    \caption{\footnotesize Left: maps of the selected boroughs of Montreal that were contaminated into outlying areas based on their offset sizes, for each simulation scenario. Right: Percentage of times each borough is detected as a potential outlier according to the four versions of the proposed model across the two simulation scenarios. A borough is defined as a potential outlier when $\kappa_u<1$, where $\kappa_u$ is the upper limit of the 95\% posterior credible interval for $\kappa$.}
    \label{fig:Sim_Out}
\end{figure}

Two simulation scenarios are considered to experiment with the prior specification of the scaling mixture components. In both scenarios, the overall log risk is $\beta_0 = -1$ and the latent effects $b_{it}, \ i=1, \dots, n, \ t=1, \dots, T$ are generated according to the Rushworth model (\ref{eq:Rush}) with $\lambda=0.7$, $\alpha=0.85$, and $\sigma = 0.3$. Both simulation examples use offsets $E_1, \dots, E_n$ taken from the analysis of COVID-19 cases shown in Section \ref{sec:Covid_Mtl}. The offsets are sorted into five categories based on their magnitude. The levels are termed ``Small" ($E < 26$), ``Medium low" ($E \in [26, 45)$), ``Medium" ($E \in [45,108)$), ``Medium high" ($E \in [108, 147)$), and ``High" ($E \geq 147$). In each simulation scenario, five boroughs (one per offset category) are then selected to be contaminated into outlying areas at given times. In one case, the selected boroughs are distant from each other and do not share a border, and in the second scenario, the five boroughs are neighbours. The maps on the left-hand side of Figure \ref{fig:Sim_Out} highlight the selected outliers based on their offset size, for each simulation scenario. For $j$ denoting one of the five selected areas in each scenario, we contaminate its latent effect as follows: $b_{jt}^{\mathrm{contaminated}} = b_{jt} + r_{jt}\times c_{jt}$, with $c_{jt} \sim \mathcal{U}\left(\max(|b_{(1)t}|, |b_{(n)t}|), 1.5\max(|b_{(1)t}|, |b_{(n)t}|)\right),$ where $b_{(1)t}$ and $b_{(n)t}$ denote the minimum and maximum generated latent effects at time $t$, respectively. The quantity $r_{jt} \in \{0,1\}$ determines whether the $j$th area is outlying at time $t$ as follows: for $t=1$, $r_{jt} \sim \mathrm{Ber}(0.4)$, and for $t \geq 2$, $r_{jt} = r_{j t-1}$ with probability $0.8$, or $r_{jt} \sim \mathrm{Ber}(0.4)$ otherwise. Figure \ref{fig:Sim_LatentGen} in Appendix \ref{App:Sim_Surplus} shows the latent effects generated from the Rushworth model before and after contamination. Finally, for each simulation scenario, $R=100$ datasets of $n=33$ boroughs and $T=52$ time points are created according to the hierarchical Poisson model $Y_{it} \sim \mathrm{Pois}(E_i\exp(\beta_0 + b_{it}))$.

Six models are fitted to the data generated for each simulation scenario. First, because \cite{Rushworth} discuss the necessity to estimate the parameter $\alpha$, the Rushworth model (\ref{eq:Rush}) is fitted with $\alpha=1$ and with a prior $\alpha \sim \mathcal{U}(-1,1)$, denoted R(1) and R($\alpha$), respectively. Then, four versions of the proposed Heavy Rushworth model (\ref{eq:HeavyRush}) are considered: two impose $\alpha=1$ and the two others assume a uniform prior, $\alpha \sim \mathcal{U}(-1,1)$. For each pair, one version, denoted HR($\cdot$),  imposes independent gamma priors to the scaling parameters, $\kappa_i \sim \mathrm{Gamma}(\nu/2,\nu/2)$, with hyperparameter as discussed in Section \ref{sec:Methods}, $\nu \sim \mathrm{Exp}(1/4)$, while the other, denoted HR-LPC($\cdot$), imposes the spatially structured prior for $\bm{\kappa}$ that is defined in Section \ref{sec:Methods} with hyperparameter $\nu \sim \mathrm{Exp}(1/0.3)$. The prior assignment and notation of the six models considered are summarised in Table \ref{tab:Models_Notation}. 

\begin{table}[H]
\footnotesize
    \centering
    \begin{tabular}{c ccc}
    \toprule
    Model & $\alpha$ & $\bm{\kappa}$ & $\nu$ \\
    \midrule
    R(1) & 1 & -- & -- \\
    R($\alpha$) & $\mathcal{U}(-1,1)$ & -- & -- \\
    HR(1) & 1 & $\mathrm{Gamma}(\nu/2, \nu/2)$ & $\mathrm{Exp}(1/4)$ \\
    HR($\alpha$) & $\mathcal{U}(-1,1)$ & $\mathrm{Gamma}(\nu/2, \nu/2)$ & $\mathrm{Exp}(1/4)$ \\
    HR-LPC(1) & 1 & log-PCAR & $\mathrm{Exp}(1/0.3)$ \\
    HR-LPC($\alpha$) & $\mathcal{U}(-1,1)$ & log-PCAR & $\mathrm{Exp}(1/0.3)$ \\
    \bottomrule
    \end{tabular}
    \caption{\footnotesize Notation and description of the six models fitted to the simulated data.}
    \label{tab:Models_Notation}
\end{table}

For each model, the MCMC procedure with two chains converged after 5,000 iterations with a burn-in period of 2,500 iterations and a thinning factor of 5, as assessed by the traceplots, effective sample sizes and $\widehat{R}$ statistic \citep{Rhat, Rhat_Stan}.

Figure \ref{fig:Sim_Out} and Table \ref{tab:Sim_SensSpe} show how often the four versions of the proposed model identify the correct set of contaminated boroughs, depending on the simulation scenario. Area $i$ is identified as an outlier when $\kappa_{u,i}<1$, where $\kappa_{u,i}$ denotes the upper limit of the posterior 95\% credible interval for $\kappa_i$. In Table \ref{tab:Sim_SensSpe}, the sensitivity measures the frequency of correct outlier identification (\%) and the specificity quantifies how often the models do not point out areas that are not contaminated (\%). For both measures, higher values are preferred. When the contaminated boroughs are not neighbours, the proposals HR($\alpha$) and HR-LPC($\alpha$), which estimate $\alpha$, perform better than the ones with fixed $\alpha=1$. In particular, HR($1$) and HR-LPC(1) correctly identify Sainte-Anne-de-Bellevue (purple borough in Figure \ref{fig:Sim_Out}) only 26\% and 5\% of the time, respectively, while HR-LPC($\alpha$) and HR($\alpha$) reach 69\% and 88\% sensitivity values for this contaminated borough with a small offset. A similar result is obtained in the second scenario, where both HR($\alpha$) and HR-LPC($\alpha$) identify Montréal-Ouest (purple borough) at least 90\% of the time, whereas the versions with fixed $\alpha=1$ do not find this borough in more than 50\% of the replicates. When the offsets are larger, in both simulation scenarios, the four versions of the proposed model accurately point out the correct outliers 100\% of the time. In terms of smoothing, in both simulation scenarios, all models equally succeed in not identifying irrelevant areas (e.g., overall specificities above 99.6). 

\begin{table}[H]
\footnotesize
\centering
\begin{tabular}{c c cccc }
  \toprule
 & Offset category & HR($1$) & HR($\alpha$)& HR-LPC($1$) & HR-LPC($\alpha$) \\ 
  \midrule
  \multicolumn{5}{l}{\textbf{Distant outliers}}\\
  \midrule
 \multirow{6}{*}{Sensitiviy} &  Small & 26.0 & 88.0 & 5.0 & 69.0 \\ 
 &   Medium low & 100.0 & 100.0 & 100.0 & 100.0 \\ 
 &   Medium & 100.0 & 100.0 & 100.0 & 100.0 \\ 
&   Medium high & 100.0 & 100.0 & 100.0 & 100.0 \\  
 &   High & 100.0 & 100.0 & 100.0 & 100.0 \\ 
& Overall & 85.2 & 97.6 & 81.0 & 93.8 \\
\addlinespace
\multirow{6}{*}{Specificity} &   Small & 99.8 & 99.8 & 99.8 & 99.8 \\ 
 &   Medium low & 100.0 & 100.0 & 100.0 & 100.0 \\
&   Medium & 100.0 & 100.0 & 99.8 & 100.0 \\ 
 &   Medium high & 99.2 & 100.0 & 100.0 & 100.0 \\ 
 &   High &  99.8 & 100.0 & 100.0 & 100.0 \\ 
& Overall & 99.8 & 99.9 & 99.9 & 99.9 \\ 
\midrule
\multicolumn{5}{l}{\textbf{Neighbouring outliers}} \\
\midrule 
 \multirow{6}{*}{Sensitiviy} &  Small & 50.0 & 90.0 & 40.0 & 91.0 \\ 
 &   Medium low &  100.0 & 100.0 & 99.0 & 100.0 \\ 
 &   Medium & 100.0 & 100.0 & 100.0 & 100.0\\ 
&   Medium high & 100.0 & 100.0 & 100.0 & 100.0 \\ 
 &   High & 100.0 & 100.0  & 100.0 & 100.0 \\ 
& Overall & 90.0 & 98.0 & 87.8 & 98.2\\
\addlinespace
\multirow{6}{*}{Specificity} &   Small & 100.0 & 99.8 & 100.0 & 99.8 \\ 
 &   Medium low &   99.8 & 99.8 & 100.0 & 100.0 \\ 
&   Medium &  99.7 & 99.8 & 99.7 & 99.8 \\ 
 &   Medium high &   99.0 & 99.6 & 100.0 & 100.0 \\ 
 &   High &  99.2 & 100.0 & 100.0 & 100.0 \\ 
& Overall &  99.6 & 99.8 & 99.9 & 99.9\\ 
 \bottomrule
\end{tabular}
\caption{\footnotesize Sensitivity and specificity of the outlier detection for each version of the proposed model in both simulation scenarios, depending on the offset size and overall.}
\label{tab:Sim_SensSpe}
\end{table}

Regarding the performances of the models, Table \ref{tab:Sim_WaicMse} summarises, for each simulation scenario, the average WAIC \citep{WAIC} computed across the 100 replicates for each model, as well as their average MSE within the contaminated and non-contaminated boroughs, and overall. For more details on these performance measures, Figure \ref{fig:Sim_WaicMse} in Appendix \ref{App:Sim_Surplus} shows the WAIC values across the 100 replicates for each model, as well as their average MSE across the different offset size categories. Smaller WAIC values are preferred and in both simulation scenarios, Table \ref{tab:Sim_WaicMse} shows that HR(1), HR($\alpha$), HR-LPC(1) and HR-LPC($\alpha$) perform better than R(1) and R($\alpha$) (e.g., when outliers are distant, R($\alpha$) yields 10,829 as the average WAIC, vs 10,757 for HR-LPC($\alpha$)). In terms of WAIC, HR($\alpha$) and HR-LPC($\alpha$) always perform better than the Rushworth models, or than the proposed models with fixed $\alpha=1$. Further, in this case where the true temporal dependence parameter is $\alpha=0.85$, when neighbouring boroughs are contaminated, the Rushworth model with unknown $\alpha$ is correctly pointed out by WAIC as the best model when compared to both versions of the proposed model where $\alpha=1$ is fixed. It can also be noted that, as expected, when distant boroughs are contaminated, HR($\alpha$) performs slightly better than HR-LPC($\alpha$), while the converse is observed with neighbouring outliers. Finally, in both scenarios, when $\alpha$ is estimated, both the Rushworth and Heavy Rushworth models (with independent and spatially structured $\bm{\kappa}$) perform better than their counterparts with fixed $\alpha=1$. This result is sensible, as the data were generated with $\alpha=0.85$.

With respect to the MSE based on fitted and observed values, regardless of the simulation scenario, the four versions of the proposed model tend to perform better than the Rushworth model among the contaminated boroughs (e.g., for neighbouring contaminated boroughs, the average MSEs are 8.0 and 40.0 for HR-LPC($\alpha$) and R($\alpha$), respectively). However, to a lesser extent, the converse is observed within the non-contaminated areas (e.g., overall MSEs of 14.4 and 7.7 for HR-LPC($\alpha$) and R($\alpha$), respectively, in the same scenario). Finally, in terms of MSE, regardless of the scenario, the proposed HR($\alpha$) and HR-LPC($\alpha$) tend to perform slightly better than HR(1) and HR-LPC(1) (e.g., for distant contaminated boroughs, overall average MSEs of 14.3 vs 17.6 for HR($\alpha$) and HR(1), respectively), which agrees with \cite{Rushworth}. 

\begin{table}[H]
\footnotesize
\centering
\begin{tabular}{ccc cccccc}
\toprule
 & & R(1) & R($\alpha$) & HR(1) & HR($\alpha$) & HR-LPC(1) & HR-LPC($\alpha$) \\ 
 \midrule
\multicolumn{5}{l}{\textbf{Distant outliers}} \\
\midrule 
\multicolumn{2}{c}{WAIC} & 10,912.2 & 10,829.1 & 10,819.2 & 10,755.5 & 10,814.0 & 10,757.4 \\ 
\multicolumn{2}{c}{$p_W$} & 647.5 & 631.3 & 568.7 & 569.4 & 573.0 & 572.2 \\ 
\addlinespace
\addlinespace
\multirow{3}{*}{MSE} & Contaminated & 44.4 & 33.9 & 7.4 & 6.6 & 8.9 & 7.1 \\
&  Not contaminated & 10.6 & 8.4 & 19.5 & 15.7 & 18.2 & 15.3 \\ 
& Overall & 15.7 & 12.2 & 17.6 & 14.3 & 16.8 & 14.0 \\ 
 \midrule
\multicolumn{5}{l}{\textbf{Neighbouring outliers}} \\
\midrule 
\multicolumn{2}{c}{WAIC} & 10,861.0 & 10,797.2 & 10,818.8 & 10,758.2 & 10,811.7 & 10,756.9 \\ 
\multicolumn{2}{c}{$p_W$} & 635.7 & 622.9 & 573.0 & 573.7 & 575.4 & 575.8 \\ 
\addlinespace
\addlinespace
\multirow{3}{*}{MSE} & Contaminated & 52.2 & 40.1 & 8.4 & 7.4 & 9.7 & 8.0 \\ 
 & Not contaminated & 9.4 & 7.7 & 18.0 & 14.6 & 17.3 & 14.4 \\ 
 & Overall & 15.9 & 12.6 & 16.5 & 13.5 & 16.2 & 13.4 \\ 
   \bottomrule
\end{tabular}
\caption{\footnotesize Average WAIC and MSE computed over the 100 replicates for each model and each simulation scenario under the different fitted models. The MSE results are distinguished between the contaminated boroughs, the non-contaminated ones, and overall.}
\label{tab:Sim_WaicMse}
\end{table}

\subsection{Analysis of COVID-19 cases in Montreal during the second wave}
\label{sec:Covid_Mtl}

The number of COVID-19 cases were recorded weekly across the $n=33$ boroughs of Montreal during the second wave, which consisted of $T=30$ weeks between August 23rd 2020 and March 20th 2021 \citep{INSPQ_CovidTimeline}. Let $Y_{it}$ be the number of cases recorded in the $i$th borough during the $t$th week, for $i=1, \dots, n$ and $t = 1, \dots, T$. Following Section \ref{sec:Methods}, the number of COVID-19 cases is modelled via a Poisson distribution whose offset $E_i$ is computed using the total number of cases and the population size of each borough, denoted $P$. We compute $E_i = \left(\sum_{i,t}Y_{it}/\sum_i P_i\right)\times P_i / T$ \citep{freitas2021spatio}. Further, three auxiliary variables that are measured at the borough level are scaled and included to model the COVID-19 cases: the median age, the percentage of the population aged 25-64 with a university diploma, and the number of beds in CHSLDs. 
Figure \ref{fig:Covid_MapsSMR_NumCases} in Section \ref{sec:Motivation} shows the SMR distribution across the boroughs in Montreal at three different points in time, alongside the evolution of the total number of cases. It can be seen that some boroughs have elevated SMRs across some weeks, but not all. For example, during the week of October 18th 2020, Mont-Royal and Dorval have higher SMRs than the rest of Montreal, as well as higher SMRs than those observed in those two boroughs during previous weeks or following ones. 
The aim of this analysis is to identify potential outlying boroughs, after accounting for the covariates' effect. Similar to Section \ref{sec:Simulation}, four versions of the proposed model are fitted to these weekly counts. The models are again denoted HR(1), HR($\alpha$), HR-LPC(1) and HR-LPC($\alpha$), as summarised in Table \ref{tab:Models_Notation}.
The performance of the proposed model is compared to that of the Rushworth model, both by fixing $\alpha=1$ and imposing a prior $\alpha \sim \mathcal{U}(-1,1)$. All prior specifications follow the same ones used in the simulation study (Section \ref{sec:Simulation}).

The six models are fitted in \texttt{R} using the \texttt{rstan} package \citep{Stan}. Convergence of two MCMC chains is attained after 10,000 iterations with a burn-in period of 5,000 and a thinning factor of 5. The diagnostics used to assess convergence are the trace plots, the effective sample sizes, and the $\widehat{R}$ statistic \citep{Rhat, Rhat_Stan}.

Table \ref{tab:Covid_Res} shows the performance measures for each model, and the estimated posterior summaries for the parameters. In terms of WAIC, smaller values are preferred, and the proposed model always performs better than the two versions of the Rushworth model (e.g., 6779 vs 6870, for HR-LPC($\alpha$) and R($\alpha$), respectively). All models that allowed the temporal dependence parameter $\alpha$ to be estimated yielded smaller WAICs than the ones that fixed $\alpha=1$. In fact, R($\alpha$), HR($\alpha$), and HR-LPC($\alpha$) resulted in 95\% posterior credible intervals for this parameter that were smaller than 1 (approximately $(0.8, 0.9)$). Finally, regarding the WAIC values, HR($\alpha$) performs the best among the four versions of the proposed model, which seems to indicate that there is no need for spatially structured scaling mixture components when analysing COVID-19 cases in Montreal during the second wave. The overall MSE results agree with the WAIC ones: the models that estimate $\alpha$ perform better than the ones with fixed $\alpha=1$ (19, 21 and 21, for R($\alpha$), HR($\alpha$) and HR-LPC($\alpha$), respectively, vs 23, 26 and 24 for their respective counterparts). The MSE results are further distinguished between the boroughs identified as potential outliers by the proposed model, and the rest of Montreal. In the potentially outlying areas, the four versions of the proposed model yield MSEs that are about 3 times smaller than the Rushworth ones. However, in the boroughs that are not found to be potential outliers by the proposed model, the MSEs that result from fitting the Rushworth models are 1.7 times smaller than the ones obtained from the proposed Heavy Rushworth models.


The three covariate effects are found to be weak by all models. The posterior 95\% credible intervals for the coefficient corresponding to the percent of the population aged 25-64 with a university diploma all include 0 (e.g., $(-1, 0.3)$ for HR($\alpha$)). While the credible interval for the age regression parameter includes 0 in the R(1) model ($(-0.3, 0.0)$), they are negative and are on the cusp of including 0 in the HR(1), HR($\alpha$) and R($\alpha$) models (e.g., $(-0.22, -0.01)$ for R($\alpha$)). The same result is obtained for the number of CHSLD beds' parameter, where the posterior credible intervals almost include 0 (e.g., (-0.18, -0.00) for HR(1) and (-0.12, 0.01) for HR($\alpha$)). 

\begin{table}[H]
{\footnotesize
	\centering
	\begin{tabular}{l c c c c c c c}
		\toprule
		& & R($1$) & R($\alpha$) & HR($1$)  & HR($\alpha$) & HR-LPC(1) & HR-LPC($\alpha$) \\
		\midrule
		\multicolumn{5}{l}{Model fit performance measures} \\
		\midrule
		\multicolumn{2}{c}{WAIC} & 6921.9 & 6870.3 & 6803.3 & 6764.6 & 6793.3 & 6778.6 \\
		\multicolumn{2}{c}{$p_W$} & 468.8  & 458.8  & 437.6 & 425.6 & 435.7  & 430.1 \\
		\addlinespace
		\addlinespace
		\multirow{3}{*}{MSE} & Outliers &  46.3 & 39.2 & 11.7 & 11.4 & 14.2 & 14.5\\
  & Not outliers & 17.9 & 15.1 & 29.4 & 23.0&  25.8 & 22.5\\
  & Overall & 23.1 & 19.5 & 26.2 & 20.9 & 23.7 & 21.1\\
		\addlinespace
  \midrule
		\multicolumn{5}{l}{Posterior summaries for the parameters: Mean (95\% CI)}\\
		\midrule
		\multicolumn{2}{c}{\multirow{2}{*}{$\beta_0$}} & -2.56  & -1.49  & -2.25  & -1.37 & -2.55  & -2.45  \\
 & & (-3.13,-1.99) & (-2.27,-0.78) & (-2.81,-1.64) & (-1.96,-0.86) & (-2.96,-2.11) & (-2.90,-2.04)\\
 \addlinespace
		\multicolumn{2}{c}{\multirow{2}{*}{$\beta_\mathrm{diploma}$}} & 0.16  & -0.46  & 0.04  & -0.40 &  0.08 & -0.08   \\
 & & (-0.68,1.01) & (-1.09,0.24) & (-0.86,0.90) & (-1.05,0.30) & (-0.79,0.89) & (-0.88,0.75) \\
  \addlinespace
  \multicolumn{2}{c}{\multirow{2}{*}{$\beta_\mathrm{age}$}} & -0.14  & -0.11  & -0.14  & -0.11 & -0.12  & -0.12  \\
 & & (-0.29,0.00) & (-0.22,-0.01) & (-0.27,-0.00) & (-0.21,-0.02) & (-0.26,0.02) & (-0.25,0.00)\\
  \addlinespace
  \multicolumn{2}{c}{\multirow{2}{*}{$\beta_\mathrm{beds}$}} & -0.11  & -0.05  & -0.09  & -0.05 & -0.09 & -0.06  \\
  && (-0.22,-0.00) & (-0.13,0.02) & (-0.18,-0.00) & (-0.12,0.01) &  (-0.20,0.00) & (-0.16,0.02) \\
   \addlinespace
  \multicolumn{2}{c}{\multirow{2}{*}{$\alpha$}} & - & 0.87  & -- & 0.87 & -- & 0.92 \\
 & & & (0.82,0.92) & & (0.81,0.93) & & (0.87,0.97)\\
  \addlinespace
		\multicolumn{2}{c}{\multirow{2}{*}{$\lambda$}} & 0.90  & 0.94  & 0.55  & 0.59 & 0.41  & 0.50  \\
 & & (0.85,0.94) & (0.91,0.97) & (0.32,0.88) & (0.37,0.89) & (0.18,0.76) & (0.25,0.83)\\
  \addlinespace
		\multicolumn{2}{c}{\multirow{2}{*}{$\sigma$}} & 0.43  & 0.45  & 0.31  & 0.33 & 0.32 & 0.34  \\
  && (0.40,0.47) & (0.42,0.49) & (0.27,0.36) & (0.29,0.38) & (0.28,0.36)  & (0.29,0.38) \\
   \addlinespace
		\multicolumn{2}{c}{\multirow{2}{*}{$\nu$}} & - & - & 3.58  & 4.14 & 2.20 & 1.88\\
 & & & & (2.03,5.93) &  (2.22,6.91) &  (1.28,3.43) & (1.02,3.06)\\
		\bottomrule
	\end{tabular}
	\caption{\footnotesize Results from the analysis of COVID-19 reported cases in Montreal during the second wave (23/08/2020 -- 20/03/2021). Model assessment (WAIC and MSE) and parameter posterior summaries: posterior mean and 95\% credible interval (CI).}
	\label{tab:Covid_Res}
 }
\end{table}

The spatial dependence parameter $\lambda$ is estimated closer to 1 by the Rushworth models than by the four versions of the proposed model (e.g., posterior means of 0.94 and 0.59, for R($\alpha$) and HR($\alpha$), respectively). This means that at each time point when modelling the latent effects, the Rushworth model gives more weight to the spatial structure compared to the proposed model. On the other hand, the four versions of the Heavy Rushworth model estimate smaller conditional standard deviations of the random effects than the Rushworth models (e.g., posterior means of 0.33 and 0.43, for HR($\alpha$) and R($\alpha$), respectively). This result is sensible as the inclusion of the scaling mixture components $\bm{\kappa}$ in the proposed model allows the variances to be raised in the boroughs that need it, without increasing the variability in the entire city of Montreal.

\begin{figure}[H]
    \centering
    \includegraphics[width=\textwidth]{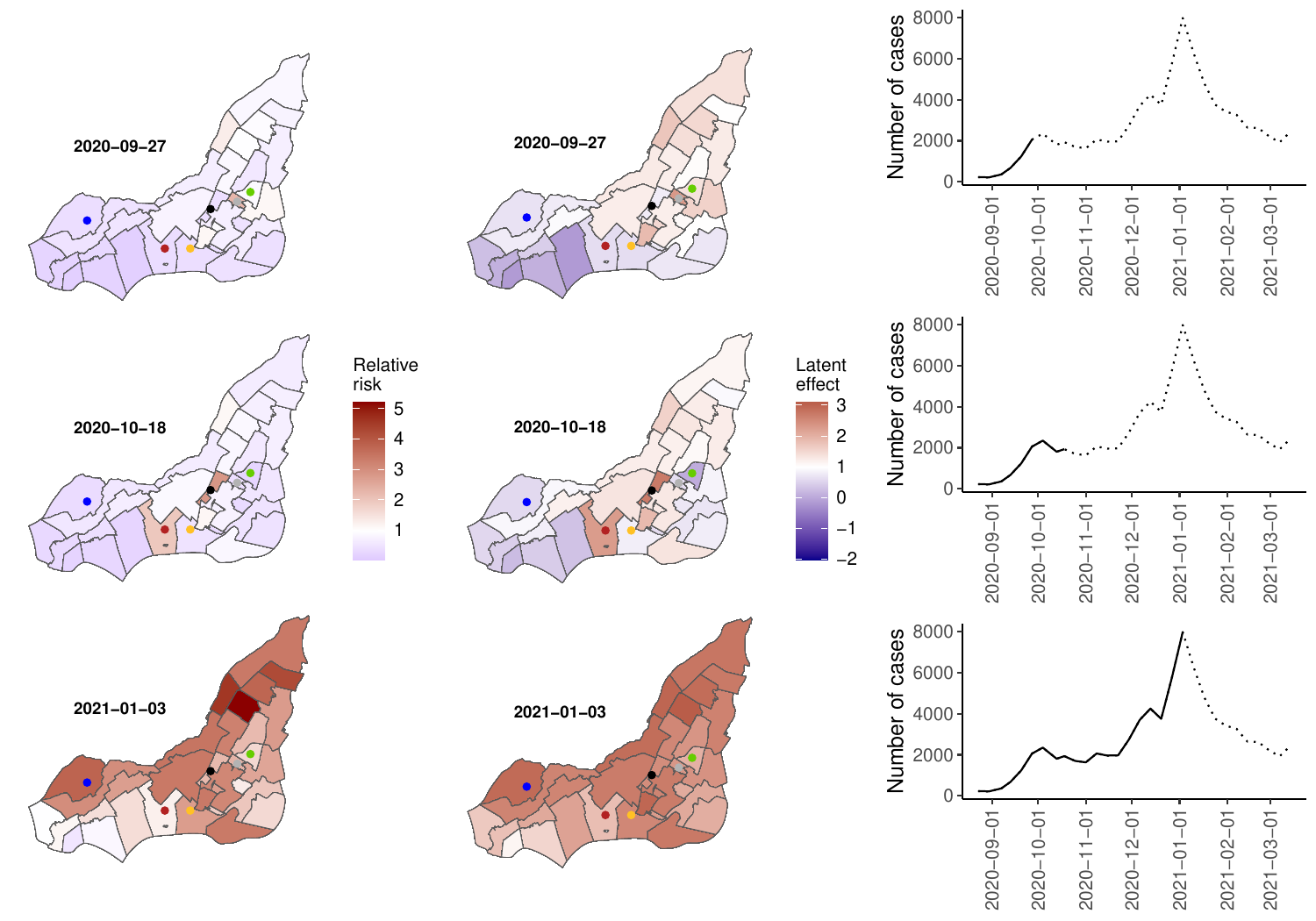}
    \caption{\footnotesize Maps of the COVID-19 relative risks (left) and latent effects (centre) estimated by the Heavy Rushworth model for three different time points across the boroughs of Montreal and total number of cases recorded over time in Montreal (right). Solid coloured circles: boroughs identified as potential outliers by the Heavy Rushworth model, the colours distinguish the boroughs to help discuss the results. The outliers are pointed out when $\kappa_u<1$, where $\kappa_u$ is the upper bound of the posterior 95\% credible interval of $\kappa$.}
    \label{fig:Covid_Res}
\end{figure}

Figure \ref{fig:Covid_Res} displays maps of the posterior means, at different time points, of the relative risks and the latent effects resulting from the proposed HR($\alpha$) model, which is the one that performed best in terms of WAIC, among all the models considered. The circles indicate which boroughs are identified as potential outliers. It is worth mentioning that the same 6 boroughs are indicated as potential outliers by the HR(1), HR-LPC(1), and HR-LPC($\alpha$) models. Similar to Section \ref{sec:Simulation}, borough $i$ is found to be a potential outlier if $\kappa_{u,i}<1$, where $\kappa_{u,i}$ is the upper limit of the 95\% posterior credible for $\kappa_i$. It is interesting to see that boroughs may be found to be potential outliers without always presenting extreme risk or latent effect. For example, Dorval (red circle) and Mont-Royal (black circle) appear to have high estimated relative risks and latent effects during the week of October 18th, 2020, compared to the other areas, without it being the case 3 weeks earlier or during the week of January 3rd, 2021, which is the second wave peak as shown on the right-hand side of Figure \ref{fig:Covid_Res}. Similarly, Outremont (gray) shows higher values during the week of September 27th, 2020, but not during the following weeks. This is the aim of the proposed model, to identify areas that may have shown unexpected behaviours at some time points, to better understand the spread of the disease and prioritise future interventions. 

It is also interesting to note that some areas might sometimes have high estimated risks without being identified as potential outliers, e.g. Saint-Leonard, the darkest borough during the week of January 3rd, 2021. The map of the posterior means of the latent effects for that week shows that after accounting for the fixed effects, the remaining latent effect for Saint-Leonard behaves like the rest of Montreal. On the other hand, le Plateau Mont-Royal (green circle), which does not appear to have extreme estimated risks over time, has an estimated latent effect that differs from its neighbours during the October 18th week. This behaviour is found to be potentially outlying by the proposed model and further investigation might help understand it.  

\subsection{Analysis of COVID-19 hospitalisations in France during the second wave}
\label{sec:Covid_Fr}

In this section, the weekly hospitalisation counts due to COVID-19 are studied across the $n=96$ French departments, during the second wave. In France, the second wave consisted of $T=26$ weeks, from July 2020 to December 2020, included \citep{INSEE_CovidTimeline}. Let $Y_{it}$ be the hospitalisation count recorded in department $i=1, \dots, 96$ at time $t=1, \dots, 26$. Similar to Section \ref{sec:Covid_Mtl}, the hospitalisations counts are modelled according to a Poisson distribution whose offsets $E$ are computed from the weekly counts and the departments' population sizes. Figure \ref{fig:Covid_MapsSMR_NumCases_Fr} in Section \ref{sec:Motivation} maps the evolution of the SMRs (top) and of the total number of hospitalisations (bottom). Some departments seem to have extreme values, compared to their neighbours or the rest of France, at some time points. For example, during the peak hospitalisation week (November 16th, 2020), Pyrénées-Atlantiques and Hautes-Pyrénées (two south-west departments) seem to have recorded higher SMRs than their neighbours, while Haute-Corse (northern Corsica) appears to be on the lower tail of the SMR distribution that week. To further study these potentially outlying behaviours, the proposed model is fitted to the data and compared to the Rushworth model. 
It is worth mentioning that since Corsica is an island, its two departments do not share a border with any other departments. There are however daily ferries travelling between these two departments and the three south-east ones on the Mediterranean Sea (Bouches-du-Rhône, Var, and Alpes-Maritimes), hence a spatial weight $w_{ij}=1$ is assigned between all of them. Additionally, the proportion of population aged 75 or older is included in the models as a covariate that is fixed through time, $x_i$. The same priors as the ones described in the previous sections are considered for the model parameters and are again denoted HR(1), HR($\alpha$), HR-LPC(1), HR-LPC($\alpha$), R(1), and R($\alpha$). 

The MCMC procedure that included two chains converged for all six models after 10,000 iterations, a burn-in period of 5,000, and a thinning factor of 5, as assessed through trace plots, effective sample sizes and $\widehat{R}$ statistics \citep{Rhat, Rhat_Stan}.

Table \ref{tab:Covid_Res_Fr} summarises the performances of the models and parameter posterior distributions. In terms of WAIC, the proposal yielded smaller values than the Rushworth model (e.g., 23,450 and 23,263 for R($\alpha$) and HR($\alpha$), respectively). The proposed HR-LPC($\alpha$) performs the best, among the six models considered, which may indicate a need for spatially structured scaling mixture components, $\kappa$'s. Note that the difference is small between the proposed models that estimate $\alpha$ and the ones with fixed $\alpha=1$ (e.g., 23,263 and 23,261 for HR(1) and HR($\alpha$), respectively). This result is sensible as both HR($\alpha$) and HR-LPC($\alpha$) estimate (0.96, 0.99) as the posterior 95\% credible interval for $\alpha$, which is very close to $\alpha=1$. On the other hand, R($\alpha$) yields a smaller WAIC value than R(1) (23,450 vs 23,468), which corresponds to $\alpha$ estimated smaller than 1, with a posterior 95\% credible interval of (0.93, 0.96).

In terms of MSE, similar to Sections \ref{sec:Simulation} and \ref{sec:Covid_Mtl}, all versions of the proposed model result in smaller values among the departments identified as potential outliers, compared to R(1) and R($\alpha$) (e.g., 8.6 vs 27.3 for HR-LPC($\alpha$) and R($\alpha$), respectively). On the other hand, R(1) and R($\alpha$) perform better in terms of MSE among the departments that are not identified as potential outliers (e.g., 12.1 vs 31.3 for R(1) and HR(1), respectively). 

Regarding the regression coefficient, all models agree on a significantly negative relationship between the proportion of the population aged 75+ and the hospitalisation counts (e.g., posterior mean of -2.8 with 95\% credible interval (-4.5, -0.9), for HR($\alpha$)), which seems counter-intuitive. This may be due to the fact that the available hospitalisation data do not include COVID-19 counts recorded in long-term medical care centres (\textit{Établissement d'hébergement pour personnes âgées dépendantes}, EHPAD), whereas EHPADs are populated by the elderly and were the epicentre of COVID-19 cases during the second wave \citep{EHPAD_Covid}. 

Similar to the results obtained for the COVID-19 cases in Montreal, the Rushworth models estimate a higher spatial dependence parameter (e.g., posterior mean of 0.93 for R($\alpha$)) than the proposed models (e.g., posterior means of 0.52 and 0.35 for HR($\alpha$) and HR-LPC($\alpha$), respectively), indicating that a higher weight is allocated to the spatial structure when the model does not accommodate for potential outliers. Finally, the variance parameter is greater in the Rushworth model than in the proposed model, with 95\% posterior credible intervals that do not overlap (e.g., (0.50, 0.54) vs (0.31, 0.36), for R($\alpha$) and HR($\alpha$), respectively).

\begin{table}[H]
{\footnotesize
	\centering
	\begin{tabular}{l c c c c c cc}
		\toprule
		&& R($1$) & R($\alpha$) & HR($1$)  & HR($\alpha$) & HR-LPC(1) & HR-LPC($\alpha$) \\
		\midrule
		\multicolumn{5}{l}{Model fit performance measures} \\
		\midrule
		\multicolumn{2}{c}{WAIC} & 23,468.5 & 23,450.5 & 23,261.5 &  23,262.8 & 23,256.6 & 23,249.1 \\
		\multicolumn{2}{c}{$p_W$} & 1249.0 & 1246.7 & 1233.9 & 1236.4 & 1225.1 & 1222.8\\
		\addlinespace
		\addlinespace
		\multirow{3}{*}{MSE} & Outliers &  27.1 & 27.3  & 8.7 & 9.0 & 8.9 & 8.6\\
   & Not outliers & 12.1 & 12.3 & 31.3 & 29.6 & 28.9 & 27.1\\
  &  Overall & 15.8 & 15.6 & 25.7 & 24.4 & 24.3 & 23.0\\
		\addlinespace
  \midrule
		\multicolumn{5}{l}{Posterior summaries for the parameters: Mean (95\% CI)}\\
		\midrule
		\multicolumn{2}{c}{\multirow{2}{*}{$\beta_0$}} & -1.47  & -1.50  & -1.52& -1.50 & -1.50 & -1.52 \\
  & & (-1.67,-1.18) & (-1.65,-1.33) &  (-1.74,-1.38) & (-1.70,-1.32) & (-1.66,-1.33) & (-1.70,-1.36)  \\
  \addlinespace
  \multicolumn{2}{c}{\multirow{2}{*}{$\beta$}} & -2.77 & -2.48  & -2.71 & -2.84 & -2.88 & -2.63   \\
  & & (-5.37,-0.75) & (-4.15,-1.03) & (-4.07,-0.63) & (-4.52,-0.96) & (-4.47,-1.40) & (-4.17,-1.07) \\
  \addlinespace
  \multicolumn{2}{c}{\multirow{2}{*}{$\alpha$}} & - & 0.94  & - & 0.98 & - & 0.98 \\
  & & & (0.93,0.96) & & (0.96,0.99) & & (0.96,0.99) \\
  \addlinespace
		\multicolumn{2}{c}{\multirow{2}{*}{$\lambda$}} & 0.91  & 0.93  & 0.51 & 0.52 & 0.34 & 0.35  \\
 &  & (0.87,0.95) & (0.90,0.96) &  (0.38,0.68) & (0.39,0.68) & (0.23,0.49) & (0.24,0.49)\\
  \addlinespace
		\multicolumn{2}{c}{\multirow{2}{*}{$\sigma$}} & 0.52  & 0.52  & 0.33  & 0.33  & 0.32 & 0.32 \\
 &  & (0.50,0.54) & (0.50,0.54) & (0.31,0.35) & (0.31,0.36) & (0.30,0.33)&  (0.30,0.34)\\
  \addlinespace
		\multicolumn{2}{c}{\multirow{2}{*}{$\nu$}} & - & - & 3.03  & 3.20 & 2.06 & 1.94  \\
& & & & (2.22,4.03) & (2.37,4.26) & (1.52,2.79) & (1.41,2.58) \\
		\bottomrule
	\end{tabular}
	\caption{\footnotesize Results from the analysis of COVID-19 hospitalisations in France during the second wave (06/07/2020 -- 03/01/2021). Model assessment (WAIC and MSE) and parameter posterior summaries: posterior mean and 95\% credible interval (CI).}
	\label{tab:Covid_Res_Fr}
 }
\end{table}

\begin{figure}[H]
    \centering
    \includegraphics[width=\textwidth]{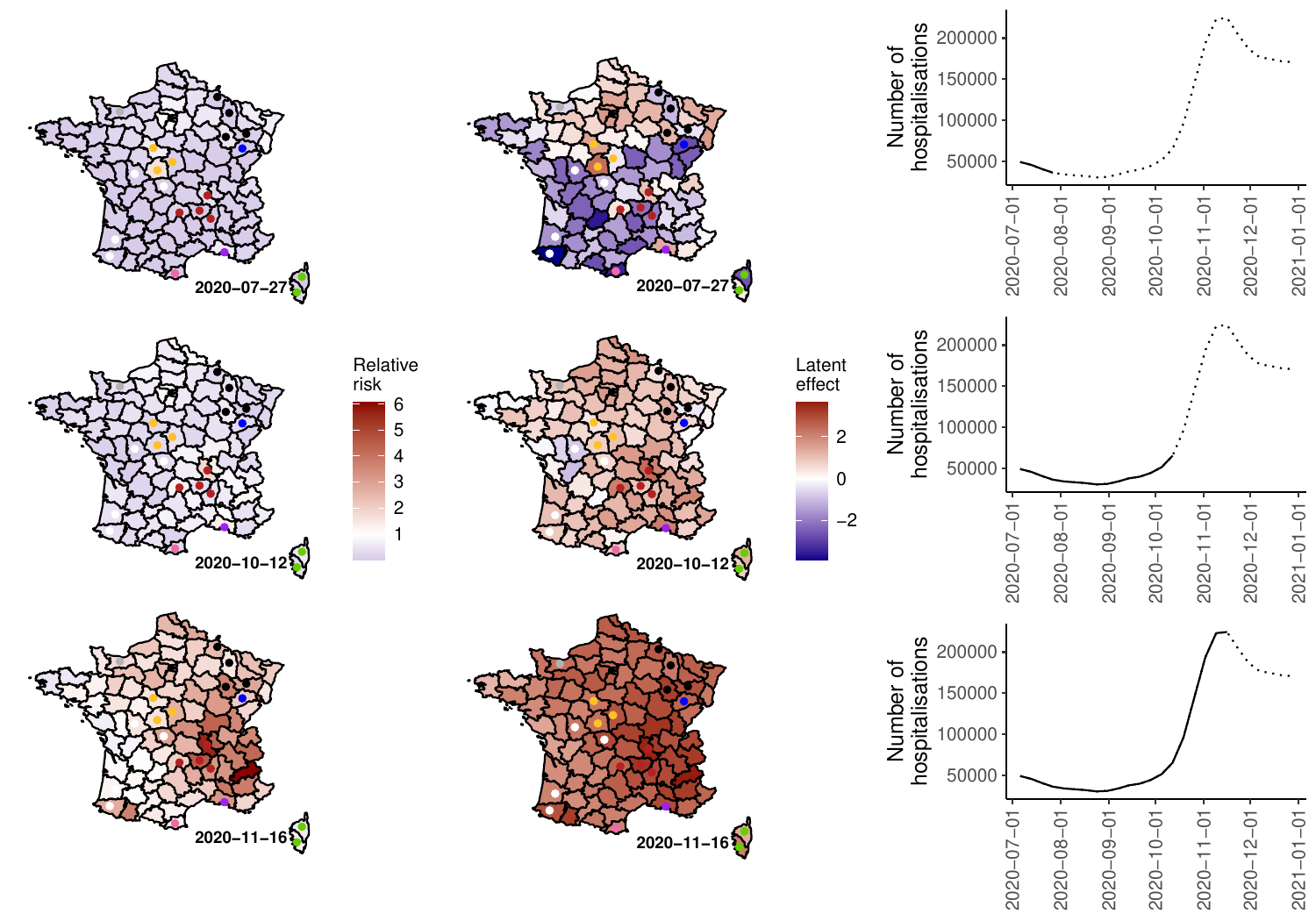}
    \caption{\footnotesize Maps of the COVID-19 relative risks (left) and latent effects (centre) estimated by the Heavy Rushworth model with spatially structured outlier indicators for three different time points across the French departments and total number of cases recorded over time in France (right). Solid coloured circles: departments identified as potential outliers by the HR-LPC($\alpha$) model, the colours correspond to the French regions to help discuss the results. The outliers are pointed out when $\kappa_u<1$, where $\kappa_u$ is the upper bound of the posterior 95\% credible interval of $\kappa$.}
    \label{fig:Covid_Res_Fr}
\end{figure}

Figure \ref{fig:Covid_Res_Fr} shows the distribution of the relative risk posterior means (left) and the latent effect posterior means (middle) at three time points, alongside the evolution of the total number of hospitalisations in France (right). The displayed posterior means correspond to the ones estimated by the proposed HR-LPC($\alpha$) model, which performed the best in terms of WAIC among the ones considered. Further, the detected potential outliers are indicated via coloured circles, where the colours correspond to different French regions, to help discuss the results. A map of the French regions in available in Figure \ref{fig:regions_Fr} in Appendix \ref{App:regions_Fr}. 
The proposed model identifies 21 departments as potential outliers, after accounting for the fixed effects. In particular, groups of neighbouring outliers seem to be identified, corresponding to different French regions. For instance, out of the 12 departments within the French region Auvergne-Rhône-Alpes (red circles), four departments are found to be potential outliers. Similarly, half of the departments in Centre-Val de Loire (yellow circles) are found to be potential outliers. Conversely, some French regions seem to gather none of the potentially outlying departments (e.g., Hauts-de-France, Bretagne). 

Similarly to the results for the analysis of COVID-19 cases in Montreal, it is interesting to note that the departments identified as potential outliers do not appear to have outlying estimated latent effects or relative risks at all time points. 
For instance, the Bouches-du-Rhône department (purple circle) is identified as a potential outlier and has a high estimated latent effect compared to its neighbours during the week of July 27th, 2020, but not later in the year. Similarly, at the beginning of the wave, northern Corsica (green circle) differs from its neighbours (e.g., Bouches-du-Rhône, due to the daily ferries) with a negative estimated latent effect. 

Finally, it can again be noted that relative risks estimated on the tails of the distribution do not necessarily coincide with an outlier identification. For example, during the wave peak, the estimated risk of hospitalisation for the Rhône department is extreme and that area is identified as a potential outlier (red circle). On the other hand, still during the wave peak, Hautes-Alpes corresponds to the other extreme estimated risk, but is not identified as a potential outlier.

\section{Discussion}
\label{sec:Discussion}

This paper proposes a spatio-temporal disease mapping model that allows for the identification of potential outliers, after accounting for fixed effects. The proposed model extends the Rushworth model \citep{Rushworth} by including a scale mixture component in the conditional variance of the spatio-temporal latent effects similarly to \cite{Congdon}.  Two prior specifications are investigated for the scaling mixture components, namely one that assumes the mixing components to follow independent gamma distributions with mean 1, and another one that allows for a spatial structure for the mixing components. It is expected \citep{Michal_zika} that the independent prior specification would perform better when the potentially outlying areas are far from each other, while situations where potential outliers are neighbours should favour the proposed model with a spatially structured prior specification. We suggest exploring both prior specifications and using WAIC to compare which model fits best. 

The results from two simulation studies help assess the ability of the proposed model to identify outlying areas in a spatio-temporal setting. The 33 boroughs of Montreal and their spatial structure were used to generate weekly data following the Rushworth model over a year (52 time points). Some areas were contaminated into being outliers. In one case, distant areas were selected, while neighbouring ones were contaminated in a second case. In both scenarios, the proposed model performed well in terms of outlier identification. In particular, when the offsets were not small, the correct areas were pointed out 100\% of the time. This result agrees with the literature on disease mapping that suggests that models perform better when the offsets are large (see, e.g., \citet{richardson2004interpreting}). The two simulation studies further showed that when outliers are neighbours, the proposed model with spatially structured scaling parameters tended to perform better than the one that assumed independent components. 

This fact was further observed in the analyses conducted on COVID-19 cases and hospitalisations in Montreal and France, respectively. In the analysis of weekly COVID-19 cases recorded across the 33 boroughs of Montreal during the second wave, the proposed model with independent scaling parameters performed better in terms of WAIC than the one assuming a spatial structure, and non-neighbouring boroughs were found to be potential outliers. On the other hand, in the analysis of weekly COVID-19 hospitalisations observed during the second wave across the French departments, the proposal with spatially structured scaling parameters performed the best among the fitted ones, and groups of neighbouring departments were pointed out as potential outliers. Finally, 6 Montreal boroughs and 21 French departments are identified as potential outliers during the second wave, after accounting for the fixed effects. Further investigation of these areas might help understand why these presented potentially outlying behaviours, when compared to the rest of the regions of interest. 

It is worth mentioning that the proposed model allows for spatial heteroscedasticity, but assumes the overall variability to be constant over time. \cite{napier2016model} extended the Leroux prior \citep{Leroux} into the spatio-temporal setting by allowing the variance parameter to vary with time. It would therefore be interesting to allow for the scaling mixture components of the proposed model to vary with space and time, similarly to the spatio-temporal dynamic linear model of \cite{fonseca2023}.

We believe that the proposed model can be useful to decision makers during an epidemic. Identifying areas that behave differently over time compared to the rest of the region may help to understand the spread of a disease better. Additionally, the proposed model may help policymakers implement localised policies or decide where to prioritise interventions. 

\subsection*{Funding}
Michal was partially supported from an award from the Fonds de Recherche Nature et Technologies [B2X - 314857], Quebec. Schmidt is grateful for
financial support from the Natural Sciences and Engineering Research Council (NSERC) of
Canada (Discovery Grant RGPIN-2017-04999) and IVADO [Fundamental Research Project, PRF-2019-6839748021].

\bibliographystyle{chicago}
\bibliography{biblio.bib}

\appendix
\counterwithin{figure}{section}
\counterwithin{table}{section}

\section{Stan code for the proposed Heavy Rushworth model}
\label{App:StanCode}

\begin{lstlisting}[caption=Stan code for the Heavy Rushworth proposed model,
  label=code]
data {
  int<lower=1> N; // number of areas
  int<lower=1> TT; // number of time points
  int<lower=1> NT; // number of areas * number of time points
  vector<lower=1, upper=N>[N] d;  // vector of the number of neighbours for each area
  matrix<lower=0, upper=1>[N,N] W; // matrix of spatial weights
  vector[N] zeros;
  int<lower=0> y[NT];  // long vector of cases ordered by time: (y_11, ..., y_n1, ..., y_1T, ..., y_nT)
  vector[NT] log_E;           // Offset in the log scale
  vector[NT] X; // vector of covariates
}

parameters {
  real beta0;            // intercept
  real beta; // regression parameter
  real<lower=0> sigma;     // conditional std deviation 
  real<lower=0, upper=1> lambda; // spatial dependence parameter
  real<lower=-1, upper=1> alpha; // temporal dependence parameter 
  vector<lower=0>[N] kappa;      // outlier indicator
  real<lower=0> nu;        // hyperparameter for kappa
  vector[NT] s;         // spatial effects
}

transformed parameters {
 matrix[N,N] PrecMat; // Precision matrix for the proposed model
 PrecMat = (1/sigma^2)*(diag_matrix(kappa .* (1-lambda + lambda*d)) - lambda * W .* (kappa*(kappa')));
}

model {
  y ~ poisson_log(log_E + beta0 + X*beta + s);
  
  // Prior for the latent effects at time 1
  s[1:N] ~ multi_normal_prec(zeros, PrecMat);
  // soft sum-to-zero constraint to avoid identifiability issues with the intercept:
  sum(s[1:N]) ~ normal(0, 0.001 * N); 
  for(t in 2:TT){
    // Prior for the latent effects at time 2, ..., T 
    s[((t-1)*N+1):((t-1)*N+N)] ~ multi_normal_prec(alpha*s[((t-2)*N+1):((t-2)*N+N)], PrecMat);
  }
  
  beta0 ~ normal(0.0, 1.0);
  beta ~ normal(0.0, 1.0);
  nu ~ exponential(1.0/4.0);
  sigma ~ normal(0.0,0.1);
  lambda ~ uniform(0.0,1.0);
  kappa ~ gamma(nu/2.0, nu/2.0);
  alpha ~ uniform(-1.0,1.0);
}
\end{lstlisting}

\section{Simulation study: data generated from the proposed model}
\label{App:SimFromModel}

In this section, we present the results from a simulation study where data are generated from the proposed model. The aim is to verify that the proposal is able to recover the true values of all the model parameters. Similar to Section \ref{sec:Simulation}, the $n=33$ boroughs of Montreal are considered over $T=52$ time points. The overall log risk is set to $\beta_0=-1$ and the offsets are taken from a Poisson distribution, $E_i \sim \mathrm{Pois}(40), \ i=1, \dots, n$. To generate the $n \times T$ latent effects $b_{it}$, the scaling mixture components $\kappa_1, \dots, \kappa_n$ are first generated from a $\mathrm{Gamma}(\nu/2, \nu/2)$ distribution, with $\nu=4$. Then, the latent effects are generated from the proposed model (\ref{eq:HeavyRush}), with $\lambda=0.9$, $\sigma=0.1$, and $\alpha=1$. Finally, $R=100$ datasets are created such that $Y_{it} \sim \mathrm{Pois}\left(E_i\exp\left(\beta_0 + b_{it}\right)\right), \ i=1, \dots, n, \ t=1, \dots, T.$ The randomness comes from generating $R$ different sets of $n\times T$ outcomes.

Again, similar to Section \ref{sec:Simulation}, the proposed model is fitted both assuming a uniform prior on $\alpha$ and fixing $\alpha=1$, denoted HR($\alpha$) and HR(1), respectively. The \texttt{rstan} \texttt{R} package is used \citep{Stan} and convergence of the two MCMC chains is attained after 5,000 iterations with a burn-in period of 2,500 and a thinning factor of 5, as assessed through trace plots, $\widehat{R}$ statistics \citep{Rhat, Rhat_Stan} and effective sample sizes.

\begin{figure}[H]
    \centering
    \includegraphics[width=\textwidth]{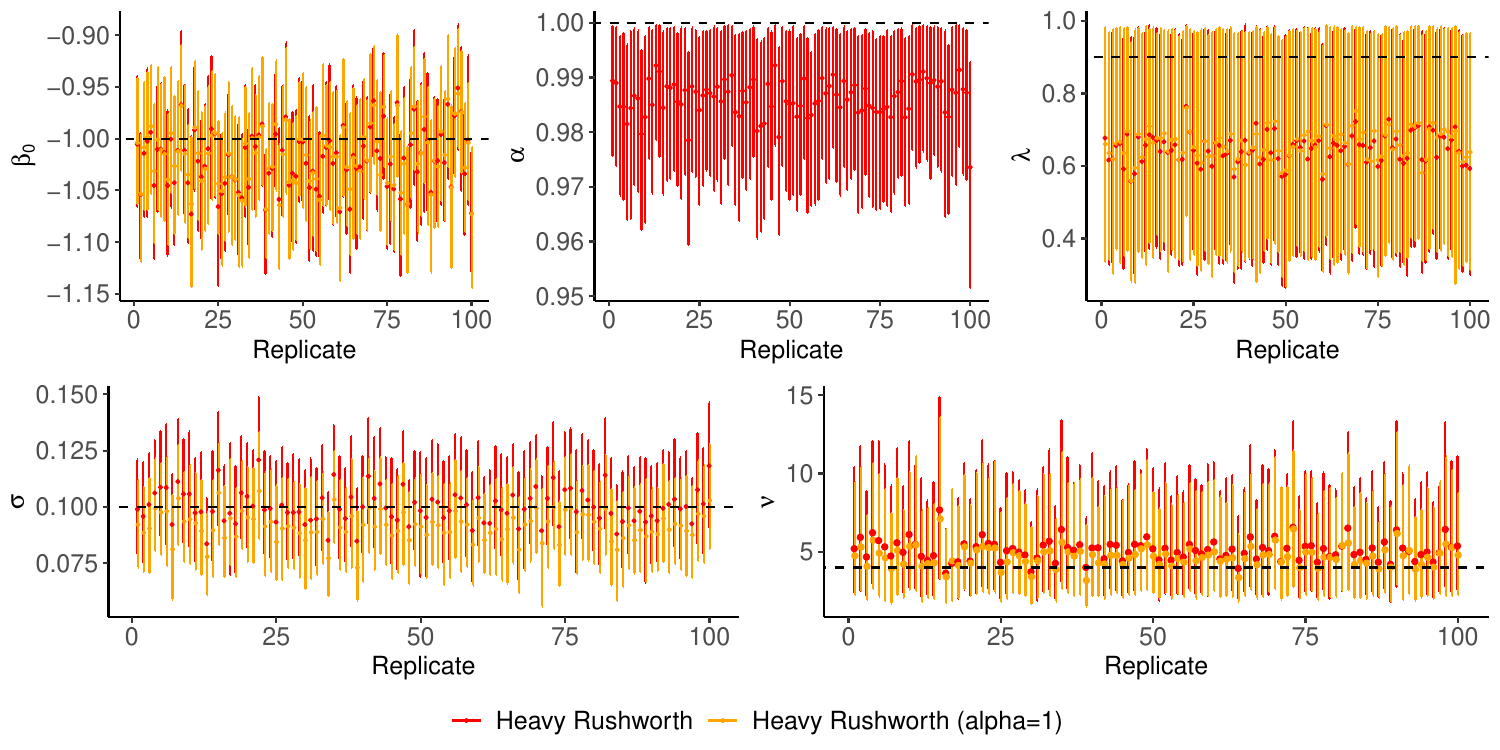}
    \caption{Parameters' posterior summaries obtained from both versions of the proposed model across the 100 replicates in the simulation study where data are generated from the proposed model. Circles: posterior means; Vertical lines: posterior 95\% credible intervals; Dashed lines: true parameter values.}
    \label{fig:SimFromModel_Param}
\end{figure}

\begin{figure}[H]
    \centering
    \includegraphics[width=\textwidth]{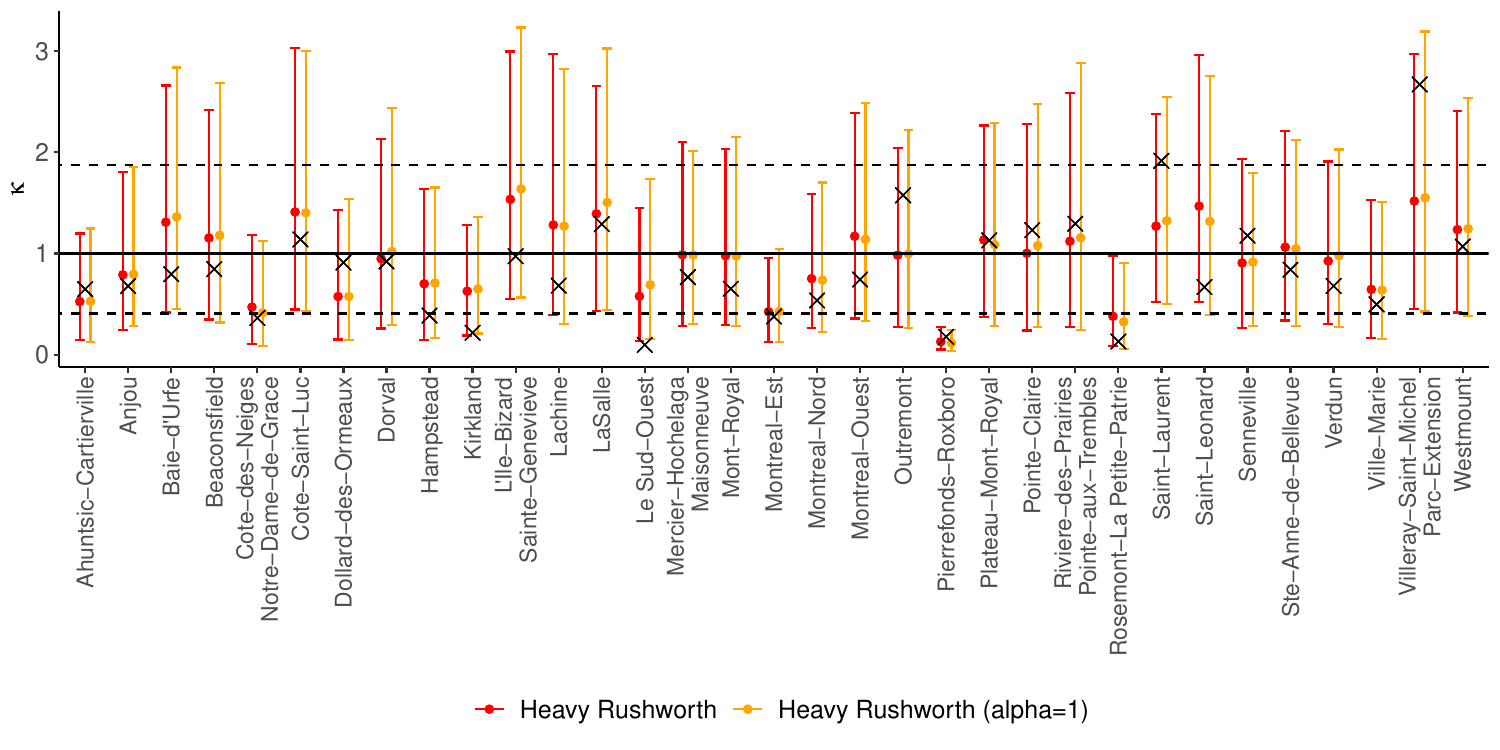}
    \caption{Posterior summaries for the scaling mixture components obtained from both versions of the proposed model in the first replicate of the simulation study where data are generated from the proposed model. Circles: posterior means; Vertical lines: posterior 95\% credible intervals; Crosses: true values; Solid horizontal line: prior mean; Dashed horizontal lines: prior 95\% credible interval.}
    \label{fig:SimFromModel_Kappa}
\end{figure}

\begin{figure}[H]
    \centering
    \includegraphics[width=\textwidth]{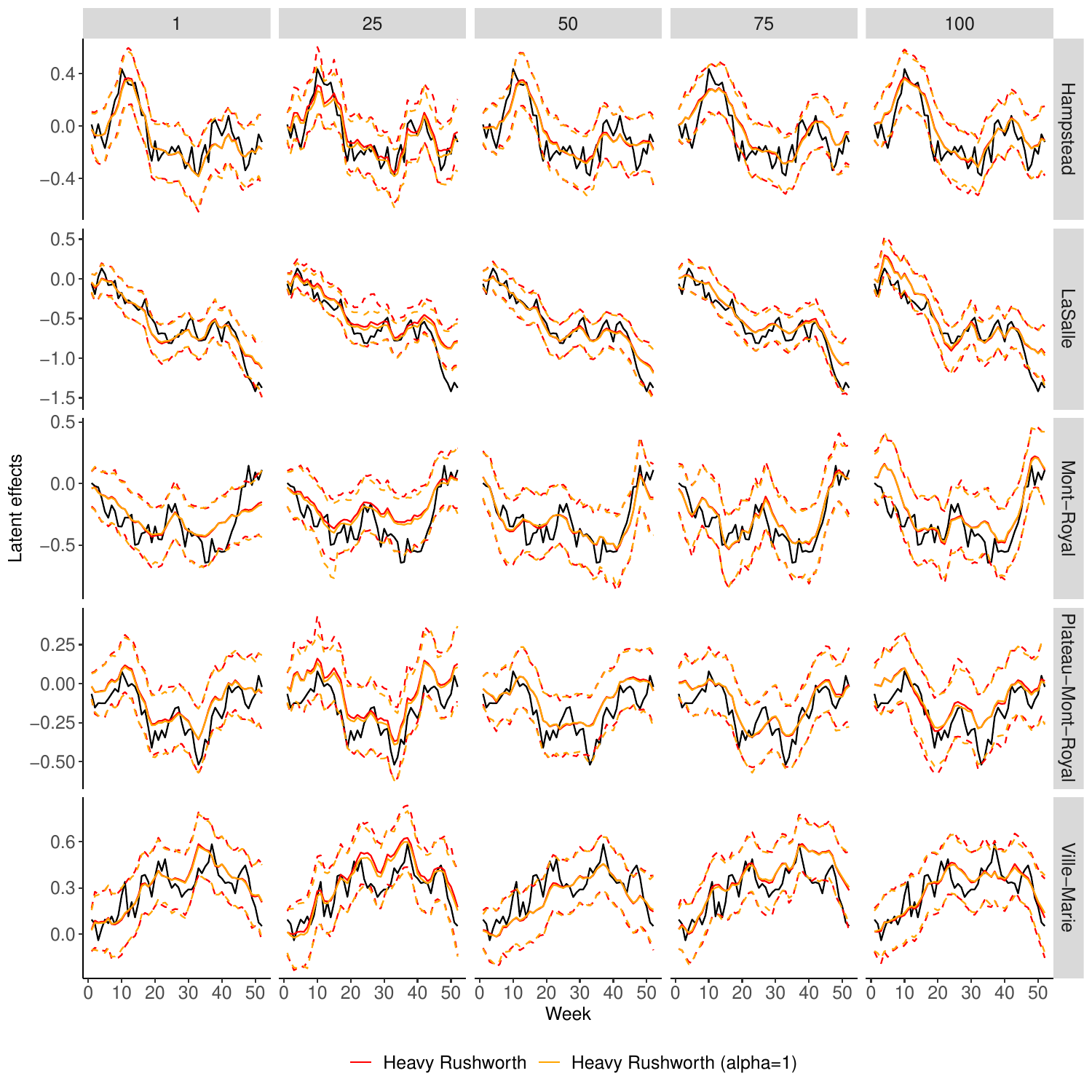}
    \caption{Posterior summaries for the latent effects obtained over time from both versions of the proposed model in 5 different boroughs (rows) across 5 different replicates (columns) of the simulation study where data are generated from the proposed model. Solid coloured lines: posterior means over time; Dashed coloured lines: 95\% posterior credible intervals; Solid black lines: true values over time.}
    \label{fig:SimFromModel_Latent}
\end{figure}

Figure \ref{fig:SimFromModel_Param} shows that across the 100 replicates, the posterior summaries (mean and 95\% credible interval) obtained from fitting both versions of the proposed model cover the true values of all the scalar parameters. It is worth mentioning that the prior for $\alpha$ is a uniform distribution over the interval (-1,1). Since this prior does not include 1, the posterior 95\% credible intervals for $\alpha$ cannot not recover 1, although the estimation is close to the truth. Interestingly, $\lambda$ tends to be underestimated on average, with posterior means approximately 0.7. However, the posterior 95\% credible intervals do recover the true values for this parameter. To a lesser extent, a similar result is observed in Appendices C and D of \cite{Michal_zika}, where data are simulated from a purely spatial disease mapping model with heavy-tailed random effects. Figure \ref{fig:SimFromModel_Kappa} shows the posterior summaries for the scaling mixture components resulting from both prior specifications in the first simulation replicate. The results for the other 99 replicates are similar to the ones presented here and across all the replicates and all boroughs. For these $\kappa$ parameters, the 95\% posterior credible intervals' coverage rates are 94.6\% and 94.5\% for HR($\alpha$) and HR(1), respectively. Further, the prior summaries (horizontal solid line and dashed lines) help visualise that the proposal is able to differ from the prior and learn from the data in order to identify potential outliers (e.g., Hampstead, Mont-Royal). Finally, Figure \ref{fig:SimFromModel_Latent} summarises the latent effects' posterior distribution estimated over time by HR(1) and HR($\alpha$) across 5 different replicates (columns) and for 5 different boroughs (rows). Through both models, the posterior means follow the true trend of the latent effects and the credible intervals capture the true values 95.5\% and 95.3\% of the time, for HR($\alpha$) and HR(1), respectively. 

\section{Supplementary material for the simulation study shown in Section \ref{sec:Simulation}}
\label{App:Sim_Surplus}

\begin{figure}[H]
    \centering
    \includegraphics[width=\textwidth]{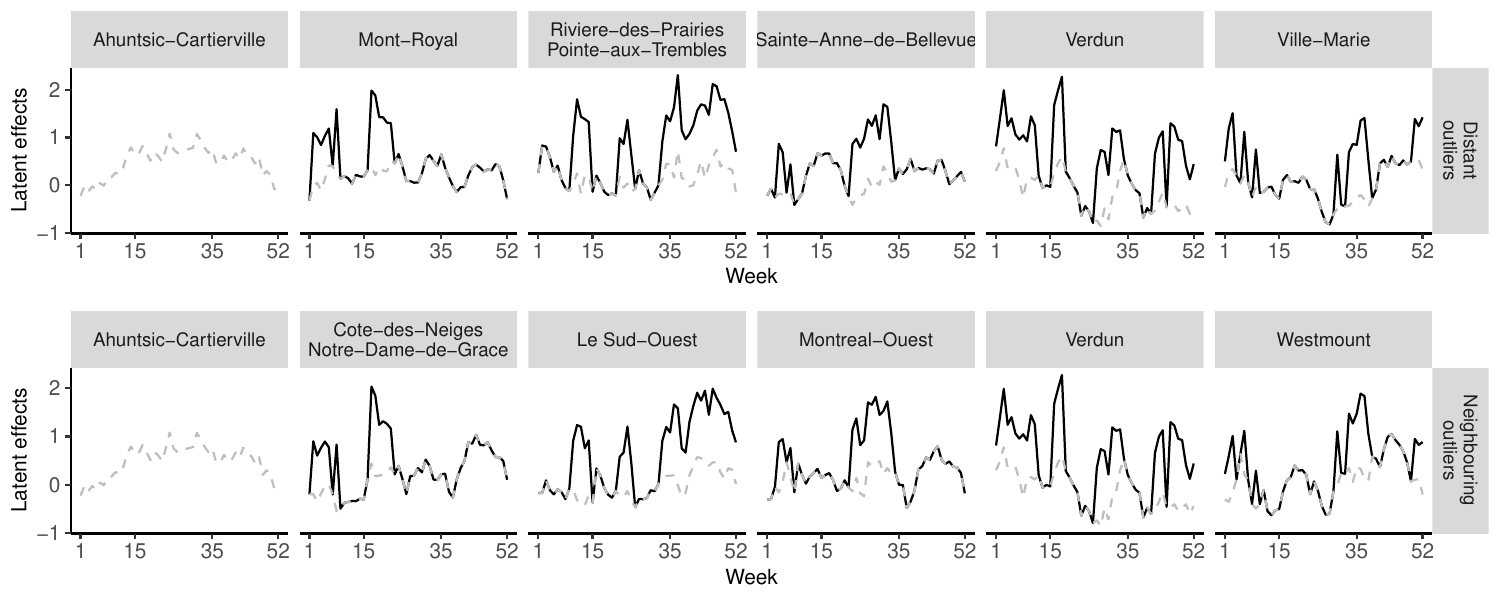}
    \caption{Generated latent effects over time for the five outliers of each simulation scenario and Ahuntsic-Cartierville. Dashed gray line: latent effects generated from the Rushworth model; Solid black line: latent effects after contamination. The periods where the two lines overlap correspond to periods of non-contamination ($r_{jt}=0$).}
    \label{fig:Sim_LatentGen}
\end{figure}

\begin{figure}[H]
    \centering
    \includegraphics[width=\textwidth]{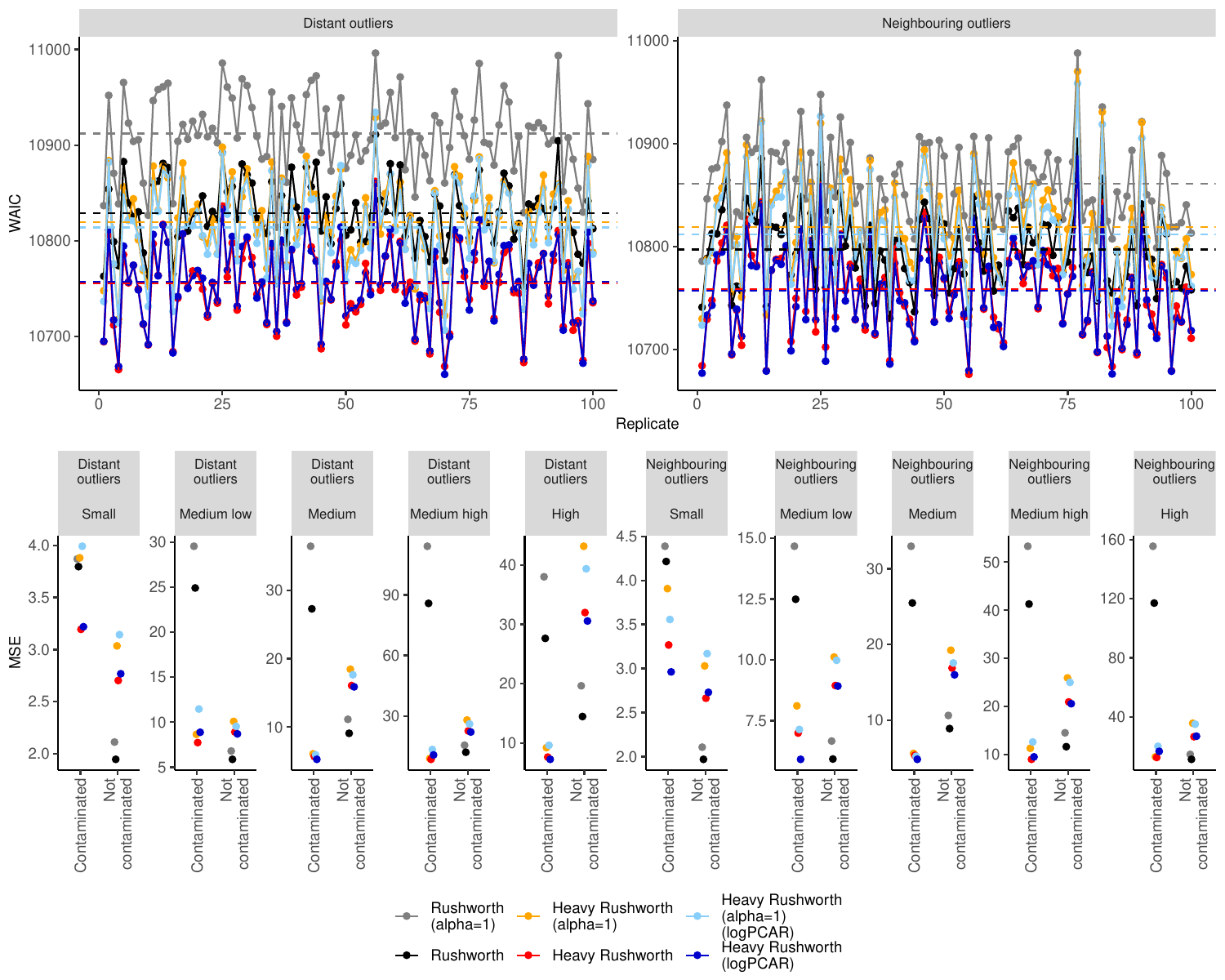}
    \caption{\footnotesize Top row: WAIC across the 100 replicates for each model and each simulation scenario under the different fitted models. Dashed lines: mean WAIC across the 100 replicates. Bottom row: Average MSE over the 100 simulation replicates for each model, offset category and scenario. The results for the contaminated boroughs are distinguished from the non-contaminated ones.}
    \label{fig:Sim_WaicMse}
\end{figure}

In this section, additional figures are presented to complete Section \ref{sec:Simulation}. Figure \ref{fig:Sim_LatentGen} shows the latent effects over time before (dashed gray line) and after (solid black line) contamination for the 5 outliers in both simulation scenarios. As a comparison, the distribution of the latent effects of a non-contaminated borough (Ahuntsic-Cartierville) is also shown over time under both simulation scenarios. 
Figure \ref{fig:Sim_WaicMse} summarises the WAIC and MSE obtained for each model and each scenario. The WAIC values are shown for each replicate and the MSEs are distinguished between the offset sizes. For each model and each replicate, let $\bm{Y}=\left[Y_{11}, \dots, Y_{n1}, \dots, Y_{1T}, \dots, Y_{nT}\right]^\top$, and the WAIC is computed as follows: $\mathrm{WAIC} = -2\sum_{i,t} \ln\left(\mathbb{E}\left[f\left(Y_{it} \mid \bm{\theta}\right) \mid \bm{Y}\right]\right) + 2\sum_{i,t} \mathbb{V}\left[\ln\left(f\left(Y_{it} \mid \bm{\theta}\right)\right)\mid \bm{Y}\right],$ where $f(\cdot \mid \bm{\theta})$ is the likelihood that corresponds to a particular model with set of parameters, $\bm{\theta}$.

\section{French regions}
\label{App:regions_Fr}

Figure \ref{fig:regions_Fr} below displays the French map where the departments are coloured according to their region. To help discuss the results from the analysis of COVID-19 hospitalisations in France during the second wave, the same colours are used here, in Figure \ref{fig:regions_Fr}, and in Figure \ref{fig:Covid_Res_Fr} in Section \ref{sec:Covid_Fr}.

\begin{figure}[H]
    \centering
    \includegraphics[width=0.7\textwidth]{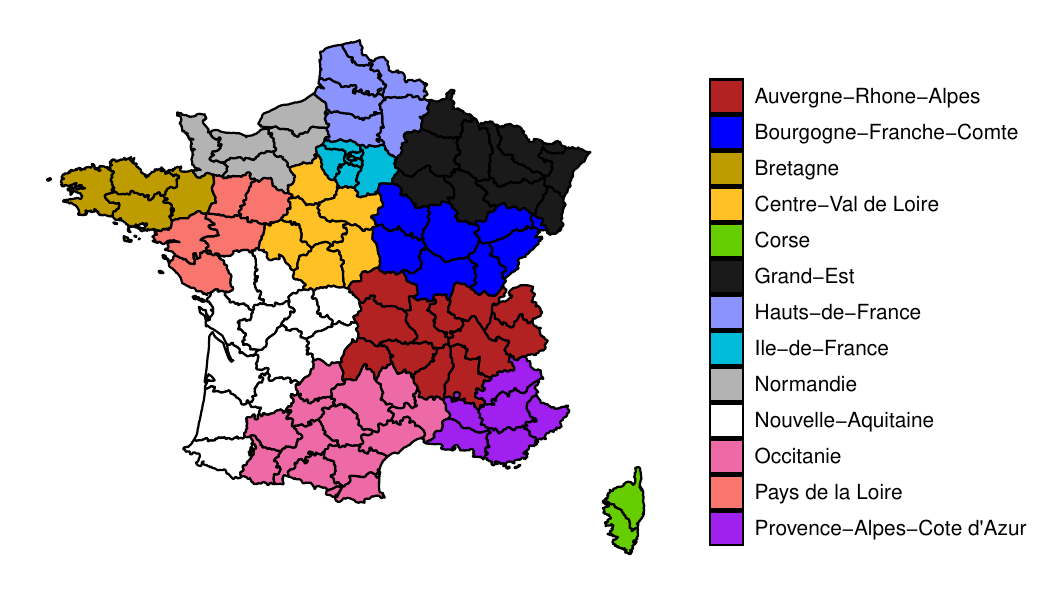}
    \caption{Map of the French regions.}
    \label{fig:regions_Fr}
\end{figure}

\end{document}